\documentclass[a4paper,11pt]{article}
\usepackage[pdftex]{graphicx}
\usepackage[T1]{fontenc}
\usepackage{lmodern}
\usepackage{slashed}
\usepackage[utf8]{inputenc}
\usepackage[english]{babel}
\usepackage{microtype}
\usepackage{cite}
\usepackage{amsmath}
\usepackage{amssymb}
\usepackage{amsfonts}
\usepackage[nottoc,notlot,notlof]{tocbibind}
\usepackage{upgreek}
\usepackage{mathtools}
\numberwithin{equation}{section}
\allowdisplaybreaks
\usepackage[all]{xy}
\usepackage{color}
\usepackage{graphicx}
\graphicspath{{images/}}
\usepackage{geometry}
\usepackage[toc,page]{appendix}
\usepackage{hyperref}
\usepackage{tikz-cd}
\usepackage{float}
\usepackage[normalem]{ulem}

\usepackage{bm}
\usepackage{ragged2e}
\usepackage{appendix}
\usepackage{slashed}
\usepackage{bbold}
\usepackage{cancel}
\usepackage{braket}

\usepackage{amsthm}

\newtheorem*{theorem*}{Theorem}

\DeclareMathAlphabet{\mathpzc}{OT1}{pzc}{m}{it}

\definecolor{blue-violet}{rgb}{0.54, 0.17, 0.89}
\definecolor{PineGreen}{cmyk}{0.92, 0, 0.59, 0.25}
\definecolor{YellowOrange}{cmyk}{0, 0.42, 1, 0}
\definecolor{orange}{rgb}{0.95, 0.5, 0.1}

\newcommand{\hash}{\mathbin{\rotatebox[origin=c]{45}{$\#$}}}

\begin{document}

\begin{titlepage}
\begin{flushright}
\par\end{flushright}
\vskip 1.5cm
\begin{center}
\textbf{\Large \bf On Superparticles and their Partition Functions }
\vskip 1cm
\vskip 0.5cm
\large {
\bf 
E.~Boffo$^{~a,}$\footnote{eugenia.boffo@fmph.uniba.sk}, P.~A.~Grassi$^{~b,}$\footnote{pietro.grassi@uniupo.it}, 
O. Hulik$^{~d,}$\footnote{ondra.hulik@gmail.com}, 
and I. Sachs$^{e}$
}  
\\
{$^{(a)}$ \it Department of Theoretical Physics, Faculty of Mathematics, Physics and Informatics,}\\
{\it Comenius University in Bratislava, Mlynsk\'{a} dolina, 842 48, Bratislava, Slovakia} \\
{$^{(b)}$ \it Dipartimento di Scienze e Innovazione Tecnologica (DiSIT),} \\
{\it Universit\`a del Piemonte Orientale, viale T.~Michel, 11, 15121 Alessandria, Italy}
\\
{$^{(c)}$ \it INFN, Sezione di Torino, via P.~Giuria 1, 10125 Torino, Italy}
\\
{$^{(d)}$ \it Theoretische Natuurkunde, Vrije Universiteit Brussel\\ Pleinlaan 2, B-1050 Brussels, Belgium}
\\
{$^{(e)}$ \it Arnold-Sommerfeld-Center for Theoretical Physics, Ludwig-Maximilians-Universit\"at M\"unchen\\ Theresienstr. 37, D-80333 Munich, Germany}

\end{center}
\vskip  0.2cm
\begin{abstract}
We describe a family of twisted partition functions for the relativistic spinning particle models. For  suitable choices of fugacities this computes a refined Euler characteristics  that counts the dimension of the physical states for arbitrary picture and, furthermore, encodes the complete BV-spectrum of the effective space-time gauge theory originating from this model upon second quantization. The relation between twisted world-line partition functions and the spectrum of the space-time theory is most easily seen on-shell but we will give an off-shell description as well. Finally we discuss the construction of a space-time action in terms of the world-line fields in analogy to string field theory.
\end{abstract}
\vfill{}
\vspace{1.5cm}
\end{titlepage}
\newpage\setcounter{footnote}{0}

\tableofcontents

\section{Introduction}
The spinning world line \cite{Howe:1988ft,Brink:1976uf,Dai:2008bh} has proved to be a valuable model for exploring some conceptual problems in the construction of off-shell string theory and string field theory in particular. Like string theory it is formulated as consistent quantization of relativistic particles but its construction is manifestly background independent and by consequence has a canonical formulation off-shell (e.g. \cite{Bastianelli:1991be,Bastianelli:1992ct,Bastianelli:2002fv,Bastianelli:2002qw,Dai:2008bh,Bonezzi:2018box,Bonezzi:2020jjq,Grigoriev:2021bes}). The world-line is rather naturally formulated in the Hamiltonian formulation reducing its analysis to that of its constraints  as well as the Becchi-Rouet-Stora-Tyutin, Batalin--Vilkovisky (BRST-BV) quantization of the latter \cite{Thorn:1988hm,Barnich:2003wj,Barnich:2004cr,Barnich:2006pc,Dai:2008bh,Bonezzi:2018box,Bonezzi:2020jjq,Bockisch:2022eas,Boffo:2022pbs,Carosi:2021wbi}. The price for this simplicity is that it misses many salient features of string theory by elimination of all the string excitations. After all, it is just an exploration of quantum field theory using the approach of string theory. On the other hand, that approach turned out nevertheless to be efficient for the analysis of 2-body problems in General Relativity, in particular for the prediction of  gravitational wave emission from such systems \cite{Jakobsen:2021smu}. 

Another distinguishing feature of string theory is  the presence of pictures \cite{Friedan:1985ge,Belopolsky:1997bg,Witten:2012bg,Castellani:2014goa,Castellani:2015paa} that naturally arise as inequivalent quantizations  of the super reparametrization invariant world sheet of  the super string. These pictures are also present in the spinning particle but have so far received little attention in the literature\footnote{See however section 4 in \cite{Dai:2008bh} some discussion.}. As in string theory the spectrum of physical states should not depend on the picture. For  the point particle, this becomes manifest by explicit inspection of the respective cohomologies. In string theory, indices or twisted partition functions are often a very efficient tool to compute the dimensions of cohomologies 
\cite{Berkovits:2005hy,Grassi:2005jz,Benvenuti:2006qr,Cremonini:2022cdm}. Here we construct an appropriate twisted partition function for the spinning world line for any picture and identify it as a generating function for the dimensions of the physical cohomologies of variable quantum numbers and picture. We will focus mostly on the spinning particle with two world line supersymmetries ($\mathcal{N}=2$ SUSY) but will make clear that our formulation holds similarly for more (or less) world line SUSY. We will furthermore show that this index encodes the dimensions of the BRST resolution of the complex into unphysical and exact subspaces. The derivation of this for on-mass shell fields is straightforward and will be given in section \ref{section3}. We then show how the same result can be obtained by an off-shell analysis in section \ref{sec:BRST-coh}. This result renders the analysis of the particle content of the  $\mathcal{N}$-extended worldline for given quantum numbers straightforward and should be helpful, in particular, for the construction  of higher spin theories on the world line.  

We then turn to the non-linear formulation of these dynamical systems. This analysis is somewhat independent of the previous part. Concretely, we describe two attempts to derive a space-time BV-action form the BRST quantization of the world-line. This turns out to be a non-trivial task. In fact, it is not even guaranteed that such a derived action exists (see e.g. \cite{Grigoriev:2021bes} for a discussion). This is in contrast to string theory where the BV-structure of the space-time action is implied by BV-master equation on the decomposition of the moduli space of Riemann surfaces \cite{Zwiebach:1993ie, Harrelson:2010zz}. The worldsheet  formulation of string theory provides a BV-morphism from the latter to the former which gives rise to the $L_\infty$ structure of the space-time action. For world-lines such a geometric structure is not known. Here we will follow two complementary approaches in section \ref{sec:action}. In the first approach we seek a product $m_2$ on the space of world line fields that is compatible with the world-sheet BRST operator $Q$. We construct such a product which is furthermore cyclic with respect to the natural inner product on the world-line, allowing us to write down an invariant world-line field theory action up to cubic order, together with the corresponding BV-transformations up to second order, reproducing the expected result for Yang-Mills theory to that order. To complete the construction one requires a 3-product, $m_3$ providing the required homotopy to cancel the non-associativity of the 2-product. We are not able to construct $m_3$ and, as explained above,  there is no a priory reason for $m_3$ to exist. We do, however, provide a quartic completion of the action that is compatible with the the full non-linear gauge invariance of Yang-Mills theory. 

In the second approach we give up the idea of realizing the field theory action in terms of multilinear products on the vector space $V$ of world line fields. Instead we consider the space-time action for the space of background fields parametrizing deformations of the BRST-charge $Q$. In this approach the space-time BV-transformations are generated by the BV-vector field $\mathfrak{s}=\frac{1}{2}\{Q,Q\}$ on the space of space-time fields with classical solutions given by the critical points of $\mathfrak{s}$  ~\cite{Horowitz:1986dta,Grigoriev:2006tt,Dai:2008bh,Bekaert:2013zya}. Resorting to the presymplectic formulation, where some of the space-time fields do not appear in the variational principle we are able to construct a cubic world-line field theory action,  with BV-symmetry induced by $\mathfrak{s}$, that includes Yang-Mills theory. We should point out that starting with the Brink-Schwarz superparticle \cite{Brink:1981nb}, where space-time -rather than world-line- supersymmetry is manifest allows to construct gauge-invariant polynomial actions on the space of functions. In particular, non-abelian super-Yang-Mills theory in 10 dimensions arises as a cubic BV-action in this way \cite{Berkovits:2001rb,Berkovits:2002zk}.

\section{Review of the Index and Partition function}

An important tool/finding of our investigation of BRST cohomology is the refined partition function with fugacities. This is a multivariable formal power series associated to a graded  vector space $ V  = \bigoplus_{I}  V _{I}$. In our context the vector space in question is the (ghost extended) Hilbert space of the spinning particle while the gradings are given by quantum numbers and ghost degree as  will be clear from the context.
As such the series takes the form 
\begin{equation}
\label{mathilb}
    \mathbb{P}_{ V } (t,r,s) =
    \sum_{n, \, p, \, k} t^n r^p s^k \, \mathrm{dim} \,  V _{n \, p}^{ k}  ={\bf tr}_{ V } 
\Big[ t^{\mathbf{n}} r^{\mathbf{p}} s^{\mathbf{k}}  \Big] \, ,
\end{equation}
with fugacities $t,r,s$ and $\mathbf{n}, \mathbf{p}, \mathbf{k}$ are the charges associated to the different gradings. Concretely, $n$ will play the role of picture number, $k$ that of a ghost degree and $p$ will measure some preserved quantum number(s) in the next section.

When the BRST charge is specified it will determine the preserved quantum numbers as well as the ghost degree (modulo linear combinations of the former) which will be raised by its action i.e. 
\begin{equation}
    Q : \,  V^k _{n \, p } \rightarrow  V^{k+1} _{n \, p}
\end{equation}
This gives rise to the \emph{twisted partition function} by specializing the fugacity $s=-1$\,,
\begin{equation}
    {\rm T} \mathbb{P}_{ V } (t,r) =
\mathbb{P}_{ V } (t,r,-1) \, .
\end{equation}
This object no longer counts the dimension of the graded components of $ V $ but rather its \emph{refined Euler characteristics} 
\begin{align}
\label{eq:E1}
    \mathbb{P}_{ V } (t,r,-1) =
    \sum_{n, \, p}^{ k} t^n r^p (-1)^k \, \mathrm{dim} \,  V _{n \, p}^{  k}
    =
\sum_{n, \, p, \, k} t^n r^p (-1)^k \,( {\rm dim}  \, Z_{n \,  p}^{ k} +
    {\rm dim} \, B_{n \, p}^{ k+1} )\nonumber
\\
    =
\sum_{n, \, p, \, k} t^n r^p (-1)^k \,(     {\rm dim}  \, B_{n \, p}^{ k} +
    {\rm dim}  \, \mathbb{H}_{n \, p}^{ k}  +
    {\rm dim} \,  B_{n \, p}^{k+1} )
    =
\sum_{n, \, p, \, k} t^n r^p (-1)^k \,
    {\rm dim}  \, \mathbb{H}_{n \,p}^{ k}
\end{align}
where we have used two exact sequences
\begin{equation}
\begin{tikzcd}
  0 \rar  & Z_{n \,  p}^{ k} \rar &  V_{n \,  p}^{ k} \rar & B_{n \,  p}^{ k+1} \rar & 0
\end{tikzcd}
\end{equation}
and
\begin{equation}
\begin{tikzcd}
  0 \rar  & B_{n \,  p}^{ k} \rar &  Z_{n \,  p}^{ k} \rar & \mathbb{H}_{n \,  p}^{ k} \rar & 0
\end{tikzcd}
\end{equation}
to infer 
\begin{gather}
    {\rm dim} \, V_{n \,  p}^{ k} = 
    {\rm dim} \, Z_{n \,  p}^{ k} +
    {\rm dim} \, B_{n \,  p}^{ k+1}\nonumber
\\
    {\rm dim} \, Z_{n \,  p}^{ k} = 
    {\rm dim} \, B_{n \,  p}^{ k} +
    {\rm dim} \, \mathbb{H}_{n \,  p}^{ k}
\end{gather}
In particular, 
\begin{equation}
    {\rm T} \mathbb{P}_{ V } (1,1) =
\chi ( V _{\bullet})
\end{equation}
reproduces the Euler characteristics. The presence of the extra fugacities allow us to read off contributions to the Euler characteristics  upon expanding $\mathbb{P}_{ V } (t,r,-1)$ in powers of $r$ and $t$ and reading off the coefficient which will give degeneracy in fixed quantum numbers $p$ and $n$ respectively. 

In the next section we will apply these techniques to a model of $\mathcal{N}=2$ (spinning) particle and its extensions. In the case that the BRST complex constitutes a resolution of the on-shell Hilbert space we will be able to read off its dimension from  $\mathbb{P}_{ V } (t,r,-1)$; for more details on this see \cite{BRST-notes}. 

In the models of our interest the Hilbert space is always freely generated by bosonic or fermionic fields on the world line with no relations among them. Under these assumptions the partition function will take the following form
\begin{equation}
    \prod_{i,j}  \frac{1}{(1-t^{n_i} r^{p_i} s^{k_i})}   (1+t^{n_j} r^{p_j} s^{k_j})  \,.   
\end{equation}
where $i$ ranges over all the bosonic fields and $j$ over the fermionic fields. 

In applications we will not be generally interested in the full partition function but rather some subsectors of given fixed quantum numbers. These can be either obtained by expanding the partition function and picking up specific powers of the fugacities or using the Molien-Weyl formula \cite{Benvenuti:2006qr}.

\section{$\mathcal{N}=2$ particle}
\label{section3}
In this section we examine the on shell spectrum of the $\mathcal{N}=2$ spinning particle via method of BRST analysis and partition functions. We start with the $\mathcal{N}=2$ model but later comment on various extensions thereof. 
 
We first examine the full spectrum of the model in various pictures in terms of the techniques based on the twisted partition function. 
We begin by recalling the world line action for the $\mathcal{N}=2$ model in the first order formalism \cite{Brink:1976uf}
\begin{equation}\label{eq:wl_action}
    S[X, P, \psi,\bar\psi,e,\chi,\bar\chi] = \int \, \mathrm{d} \tau \,  P_m \dot{X}^m - \mathrm{i} \bar\psi \dot{\psi} - \mathrm{i} \psi \dot{\bar\psi} -  \tfrac{e}{2} P^2- \mathrm{i} \chi \bar\psi^m P_m - \mathrm{i} \bar\chi \psi^m P_m\,,
\end{equation}
with Lagrange multipliers $e,\chi, \bar{\chi}$ enforcing the first class constraints that generate the $\mathcal{N}=2$ superalgebra on the worldline. Upon BRST quantization \eqref{eq:wl_action} gives rise to a graded algebra $\mathcal{A}$ generated by $(X^m, \psi^m, c, \gamma, \beta)$ and their duals
$(P_m, \bar\psi_m, b, \bar\gamma, \bar\beta)$ respectively with canonical commutation relations
\begin{eqnarray}
\label{crA}
[P_m, X^n] = -i\delta_m^n\,, ~~~~
\{\bar\psi_m, \psi^n\} = \delta_m^n\,, ~~~~
\{c, b\} = 1\,, ~~~~
[\gamma, \bar\beta] = 1\,, ~~~~
[\bar\gamma, \beta] = 1\,, 
\end{eqnarray}
The BRST charge, governing the constraints of the system, is given by 
\begin{equation}
    Q= -c P^2 + \bar{\gamma} q + \gamma \bar{q} + b \gamma \bar{\gamma} \, .
\end{equation}
In addition to the Hamiltonian constraint $P^2=0$, there is another pair of constraints 
\begin{equation}
q := \psi \cdot P \,, \quad \bar q : = \bar\psi \cdot P \,,
\end{equation}
that generate the super-reparametrizations. The BRST charge
commutes with the ${\rm }U(1)$ R charge
\begin{equation}\label{eq:R-charge}
    R = \psi \cdot \bar\psi - \gamma \bar\beta + \beta \bar\gamma 
\end{equation}

\subsection{Picture zero}\label{sec:p0}
In this section we study a representation of the algebra $\mathcal{A}$ on a Fock space build on a highest weight state $\ket{k}$ given by an on-shell ($k^2=0$) $4$-momentum $k$. 
An alternative analysis will be performed in section \ref{sec:BRST-coh}. The Weyl algebra of the super reparametrization ghosts $\gamma,\beta$ has 4 inequivalent representations labeled by pictures \cite{Friedan:1985ge, Belopolsky:1997bg,Witten:2012bg,Witten:2012bh}. In picture $0$ the vacuum state of the Fock space $ V $, generated by power series in $\{c,\gamma,\beta,\psi^m\}$  is characterized by
\begin{eqnarray}
\label{App2A}
\bar\psi_m| k\rangle =0\, 
\quad
b| k\rangle =0\,
\quad
\bar\gamma| k\rangle =0\,
\quad
\bar\beta| k\rangle =0\,,
\end{eqnarray}
on which the dual operators are realized as partial  derivatives  
\begin{equation}
\bar\psi_m = \partial_{\psi^m}
\quad
b = \partial_{c}
\quad
\bar\gamma = \partial_{\beta}
\quad
\bar\beta = \partial_{\gamma}\,,
\end{equation}
and the BRST charge of the $\mathcal{N}=2$ particle is then represented by
\begin{equation}
    Q= \partial_\beta \psi^m \partial_m + \gamma \partial_{\psi_m} \partial^m + \gamma \partial_\beta \partial_c  \, ,
\end{equation}
Thus, the massless spectrum ($k^2=0$) is subject to the constraints\begin{eqnarray}
    \label{AA1}
    \left( \gamma \partial_{\psi} \cdot k + \psi\cdot k \partial_{\beta} - \gamma  \partial_{\beta}  \partial_{c} \right)\ket{\phi}=0\, .
\end{eqnarray}
In particular, we trivialized the Hamiltonian constraint. Furthermore for a fixed (but unspecified) 4-momentum $k$, the unphysical space $ V / Z$ is finite dimensional. We will describe the quantization in the other pictures below in section \ref{sec:otherpic}.

Let us now consider the twisted partition function reviewed in the previous section. In picture zero the variable $t$ will not appear here. We start with the partition function with a maximal number of independent fugacities, 
that is, one for each independent generator
\begin{align}\label{cippaBB}
  \mathbb{P}_ V (
  \gamma, \beta, c, \psi)  
  = {\bf tr}_{ V }[ \gamma^{\mathbf{n}_\gamma}\beta^{\mathbf{n}_\beta}c^{\mathbf{n}_c}\psi^{\mathbf{n}_\psi} ]=
  \frac{(1+\psi)^D(1+c)}{(1-\gamma)(1-\beta)}
  \, .
\end{align} 
Note that in this equation we have used the same letters for the generators $\gamma,\beta,c,\psi$ and the fugacities for the corresponding number operators. This notational confusion will then be resolved in eqn \eqref{cippaBBB}. Rather than $Q$ we first consider the differential $\bar\gamma q=\bar\gamma\psi\cdot P$, with the conserved quantum numbers 
\begin{align}\label{eq:pi}
\mathbf{p}_1&=\mathbf{n}_\beta+\mathbf{n}_\psi+\mathbf{n}_\gamma \,,\quad
      \mathbf{p}_2= \mathbf{n}_c\,,\quad
      \mathbf{p}_3=\mathbf{n}_\gamma 
\end{align}
and ghost degree $ \mathbf{k}=\mathbf{n}_\gamma-\mathbf{n}_\beta+\mathbf{n}_c$. This choice for $\mathbf{k}$ is clearly not unique. For instance, $ \mathbf{k}=-\mathbf{n}_\beta$ would have been a valid definition of the ghost degree. Our choice turns out to be convenient below. Note also  that $p_3\leq p_1$. Then, 
\begin{align}\label{cippaBBB}
  \mathbb{P}_ V (
  r, v, z,s)  
 &= {\bf tr}_{ V }[ \gamma^{\mathbf{p}_3}\beta^{\mathbf{p}_2+\mathbf{p}_3-\mathbf{k}} c^{\mathbf{p}_2 }\psi^{\mathbf{p}_1+\mathbf{k}-\mathbf{p}_2-2\mathbf{p}_3} ]\nonumber\\
 &={\bf tr}_{ V }[ (\frac{\psi}{\beta})^{\mathbf{k}}(\frac{\beta\,c}{\psi})^{\mathbf{p}_2}(\frac{\gamma \,\beta}{\psi^2})^{\mathbf{p}_3}
 \psi^{\mathbf{p}_1} 
  ]\nonumber\\
  &=:{\bf tr}_{ V }[s^{\mathbf{k}}v^{\mathbf{p}_2}z^{\mathbf{p}_3}r^{\mathbf{p}_1}
  ]\nonumber\\
&=\frac{(1+r)^D(1+ sv)}{(1- szr)(1-\frac{r}{ s})} \, .
 \end{align}
It is clear from \eqref{eq:pi} that, for fixed $p_i$, $k$ is not uniquely determined. In fact this holds for any choice for the $p_i$'s and $k$.  Thus, a priori, we should expect that cohomologies at different ghost number contribute to the refined Euler characteristics for fixed $p_1$,  $p_2$ and  $p_3$. On the other hand, expanding $\mathbb{P}_V$ for $s=-1$ we have 
\begin{align}
  {\rm T} \mathbb{P}_{ V } (r,z,v)=1+r
   \left((D-1)-z\right)+r^2 \left(\frac{1}{2} (D-2) (D-1)+(1-D)
   z+z^2\right)+O\left(r^3\right)+O\left(v^1\right)\,,
\end{align}
therefore we see that the $z$-independent part at fixed power in $r$  counts the correct dimension of the de-Rham differential $p_1$-forms in $D$ dimensions. To understand how this comes about, let us list the cohomology $\mathbb{H}^\bullet_{\bar\gamma q}$ which is generated by 
\begin{gather}\label{eq:cog}
\{\gamma^m\epsilon^{(n)}\cdot\psi^{n},\gamma^m\beta^n\varepsilon^{(p)}\cdot\psi^{p},
c\gamma^m\epsilon^{(n)}\cdot\psi^{p},c \gamma^m\beta^n\varepsilon^{(p)}\cdot\psi^{ p}| \epsilon^{(n)}\sim\epsilon^{(n)}+k\wedge\epsilon^{(n-1)};
\\
 k\wedge \varepsilon^{(p)}=0; \varepsilon^{(p)}\sim\varepsilon^{(p)}+k\wedge\varepsilon^{(p-1)} \} \,.\nonumber
\end{gather}
We then find that 
\begin{align}
    \mathbb{H}_{\bar\gamma q}^k(V_{p_1,p_3})=\emptyset\,,
\end{align}
except for $k=p_3$, so that in \eqref{mathilb} only one term contributes in the sum over $k$. As a consequence, the refined Euler characteristics, in fact, counts the dimension of $\mathbb{H}_{\bar\gamma q}^{p_3}(V_{p_1,p_3})$. We may repeat this exercise for $\gamma\bar q=\gamma\bar\psi\cdot P$, with conserved quantum numbers as 
\begin{align}
    \tilde {\mathbf{p}}_1&=\mathbf{n}_\beta+\mathbf{n}_\psi+\mathbf{n}_\gamma \,,\quad
     \tilde{\mathbf{p}}_2= \mathbf{n}_c,\quad
     \tilde{\mathbf{ p}}_3=\mathbf{n}_\beta \,.
\end{align}
For $s=-1$ the form of the partition function is identical with that form $\bar\gamma q$ but the power of $z$ has a different interpretation.

More interesting for our purpose is the combination $d:= \bar\gamma q+\gamma\bar q$. Since $\bar\gamma q$ and $\gamma\bar q$ (anti-) commute on shell it should hold that 
\begin{align}
    \mathbb{H}_{d}( V )= \mathbb{H}_{\gamma\bar q} (\mathbb{H}_{\bar\gamma q}( V ))\,.
\end{align}
Since $\bar\gamma q$ and $\gamma\bar q$ are simultaneously defined on the subspace with  fixed $p_1$ (but variable $p_3$), we should furthermore have the refined property
\begin{align}
     \mathbb{H}_d ( V_{p_1})=  \mathbb{H}_{\gamma\bar q} ( \mathbb{H}^{\bullet}_{\bar\gamma q }(V_{p_1}))\,,
\end{align}
where 
\begin{align}
\mathbb{H}^{\bullet}_{\bar\gamma q\;p_1}=\oplus_{p_3=0}^{p_1} \mathbb{H}^{p_3}_{\bar\gamma q\;p_1p_3}\,.
\end{align}
Now, on $\mathbb{R}^n$, $\omega\in \mathbb{H}^{\bullet}_{\bar\gamma q\;p_1}$ is $\gamma\bar q$-closed if it is $\gamma\bar q$-exact\footnote{For $\omega=\gamma^{p_3}A^{[p_1-p_3]}\cdot \psi^{(p_1-p_3)}$, $\gamma\bar q$-closure implies deRham co-closure and thus deRham co-exactness also within $\mathbb{H}^{\bullet}_{\bar\gamma q\;p_1}$, since for an exact form  $\delta\mathrm{d} A^{[n]}=-\mathrm{d}\delta A^{[p+1]}$ holds on-shell, i.e. $\delta$ is a chain map.}, except for $p_3=0$. On the other hand, for the twisted partition functions \eqref{cippaBBB}, summing over $p_3$ amounts to setting $z=1$. Then,  the refined Euler characteristics 
\begin{align}\label{eq:reE-1}
  {\rm T} \mathbb{P}_ V (
   r, 1, v)  
 &=  \sum_{p_2}v^{p_2}\sum_{p_1}r^{p_1} \sum_{p_3=0}^{p_1} (-1)^{p_3}  {\bf tr}_{\mathbb{H}_{\bar\gamma q\;p_1p_2p_3}^{p_3+p_2}}[1]   = \sum_{p_2} v^{p_2} \sum_{p_1} r^{p_1} \chi\left( V_{p_1,p_2}\right)\, 
\end{align}
localises at $p_3=0$ and thus actually counts the dimension of $\mathbb{H}^{p_2}_{\bar\gamma q\; p_1p_2\,p_3=0}$. 

Finally, we consider the differential $Q=d+b\gamma\bar\gamma$. Now, $b\gamma\bar\gamma$ (anti-)commutes with $d$ but, unlike $d$ does not preserve $p_2$. Thus 
\begin{align}
    \mathbb{H}_Q ( V_{p_1})=  \mathbb{H}_{b\gamma\bar\gamma} ( \mathbb{H}^{\bullet}_{\bar\gamma q\;p_1})\,,
\end{align}
where 
\begin{align}
\mathbb{H}^{\bullet}_{\bar\gamma  q\;p_1}=\oplus_{p_2}\mathbb{H}^{p_2}_{\bar\gamma q\;p_1p_2\,p_3=0}
\end{align}
Now, $\mathbb{H}^{\bullet}_{\gamma\bar q\;p_1}$ is in the kernel of $b\gamma\bar\gamma$ (no dependence on $\beta$) and, at the same time has empty intersection with the image of $b\gamma\bar\gamma$ (being independent of $\gamma$ since $p_3=0$). Thus, 
\begin{align}
    \mathbb{H}_Q (  V _{p_1})=\mathbb{H}^{0}_{\gamma\bar q\;p_1\,0\,0}(  V _{p_1})\oplus \mathbb{H}^{1}_{\gamma\bar q\;p_1\,1
    \,0}(  V _{p_1})\,.
    \label{full-coh}
\end{align}
The only difference between $ \mathbb{H}_d ( V_{p_1})$ and  $\mathbb{H}_Q ( V_{p_1})$ is that the former can be further restricted to a subspace of fixed $p_2$ while the latter can not\footnote{The remaining quantum number $p_1$ can the be identified with $R$ in the previous section.}. This implies that the twisted partition function $ \mathbb{P}_ V 
  (r,1, 1,-1) $ vanishes, which is reflected in 
  \eqref{cippaBBB} by the simple zero at $v=-s=1$. We can still extract the dimension of $\mathbb{H}^0_Q(V_{p_1})$ at fixed $p_2$ by simply dividing \eqref{cippaBBB} by $1+sv$ before taking the limit. Equivalently 
\begin{align}\label{eq:partQ}
     -\partial_v {\rm T} \mathbb{P}_ V 
  (r,1 , \, v)|_{v= 1}
\end{align}
counts the dimension of $\mathbb{H}^0_Q(V^{p_1})$. 

It is also instructive to display 
\begin{align}\label{ppA}
\mathbb{P}_{\Omega^{(\bullet|0)}}( r,s)\equiv \mathbb{P}_V( r, 1, 1,s) = \frac{(1 +  r)^D(1 + s)}{(1 - sr  )(1 - \tfrac{ r}{s})} \, ,
\end{align}
where we defined $\Omega^{(\bullet|0)}\equiv V$ in order to distinguish between other pictures below. $\mathbb{P}_{\Omega^{(\bullet|0)}}( r,s)$ then expands into 
  \begin{align}
  \left(1+r \left(D+s+\frac{1}{s}\right)+r^2
   \left(\frac{D(D-1)}{2}+1+D
   s +s^2+\frac{D}{s}+\frac{1}{s^2}\right)+O\left(r^3\right)\right)(1+s)
\end{align}
making the ghost and ghost-for-ghost contribution to the spectrum of 1- and 2-forms Maxwell fields manifest. For example in the $r^1$ subsector we have the contribution 
\begin{equation}
 (1 + s)r (D + s  + \tfrac{1}{s})
 \end{equation}
which represents the helicity  spin one degrees of freedom $D + s + 1/s$ ($D$ bosonic degrees  of freedom  of the gauge potential and $s, 1/s$ represent the ghost fields) together with  the respective antifields, accounted for by the factor $(1+s)$. Alternatively, note that 
the coefficient linear in $r$ can be written as $\frac{1}s + (D+1)  + (D+1) s + s^2$, representing the target space ghost $C$, the 
gauge potential $A_m$ plus an auxiliary field $B$, the antifields $A^*_m, B^*$ and the antifield of the ghost field $C^*$. 
To conclude we wish to stress again that the quantum number $R:=p_1$  as defined in  \eqref{eq:pi} is the $R$-charge \eqref{eq:R-charge} of the $\mathcal{N}=2$ world line SUSY algebra. 

\subsection{Other pictures and the full partition function}\label{sec:otherpic}

The remaining inequivalent representations of the Weyl algebra of the $\beta$- and $\gamma$ ghosts are obtained by replacing powers of the latter by derivatives  of their $\delta$-function (Dirac delta) distributions, 
 \begin{align}
 \partial^{n}_\beta\delta(\beta)\quad \text{and}\quad  \partial^{m}_\gamma\delta(\gamma)\,.
 \end{align}
 We will denote different picture sectors of the Hilbert space as
$\Omega^{(\bullet|0)},\Omega^{(\bullet|1)}_\gamma, \Omega^{(\bullet|1)}_\beta, \Omega^{(\bullet|2)}$. 
To simplify the computation of the partition function we will drop $p_2$ and $p_3$, and focus on $\mathbb{P} (r,1,1,s)$. Later considerations on the cohomology groups will rely on the existence of isomorphisms such as the Hodge operator or the \emph{picture changing operators}.
For convenience we enumerate in table \ref{redefined q.n.} all the charges of the various generators and states appearing in the $\mathcal{N}=2$ model, 
\begin{table}
\begin{center}
\begin{tabular}{|c|c|c|c|c|c|c|c|c|c|c|c|c|c|c|}
\hline
Field & $X^m$ & $\psi^m$ & $c$ & $\gamma$ & $\beta$ & 
$P_m$ & $\bar\psi^m$ & $b$ & $\bar\gamma$ & $\bar\beta$&  $ \delta(\gamma)$ & $\delta(\beta)$ &  $\delta(\bar\gamma)$ & $\delta(\bar\beta)$\nonumber \\
\hline
$r$
& 0 & 1 & 0 & 1 & 1  
& 0 & --1 & 0 & --1 & --1 & -1& -1 & 1 & 1 
\\
\hline
$s$ 
& 0 & 0 & +1 & +1 & --1  
& 0 & 0 & --1 & +1 & --1 & -1 & 1 & -1 & 1   \\
\hline
$\Pi_{r,s}$ & 1 & $r$ & $s$ & $sr$ & $\frac{r}{s}$ & $1$ & $\frac{1}{r}$ & $\frac{1}{s}$ & $\frac{s}{r}$ & $\frac{1}{sr}$&$\frac{1}{sr}$ & $\frac{s}{r}$ & $\frac{r}{s}$ & $sr$ \\
 \hline
\end{tabular}
\caption{Fugacity $r$ and ghost number $s$ for the fields.}
\label{redefined q.n.}
\end{center}
\end{table}
where the last row lists the contribution of the single word to the twisted  partition function of a given field.

The full partition function including all pictures is
\begin{align}\label{eq:fullp}
 \mathbb{P}_{\Omega^{(\bullet|\bullet)}}
 =& \, 
  \mathbb{P}_{\Omega^{(\bullet|0)}}
  +
  t\mathbb{P}_{\Omega_{\gamma}^{(\bullet|1)}}
   +
  t  \mathbb{P}_{\Omega_{\beta}^{(\bullet|1)}}
    +
   t^2 \mathbb{P}_{\Omega^{(\bullet|2)}} =\nonumber
     \\ 
= & \, \frac{(1 +  r)^D(1 +s)}{(1 - sr  )(1 - \tfrac{ r}{s})}
+
t \frac{(1 +  r)^D(1 +s)}{(1 - \tfrac{1}{sr}  )(1 - \tfrac{ r}{s})} \tfrac{1}{sr} +\nonumber
\\
\, & + t \frac{(1 +  r)^D(1 +s)}{(1 - sr  )(1 - \tfrac{s}{ r})} \tfrac{s}{ r}
+
t^2 \frac{(1 +  r)^D(1 +s)}{(1 - \tfrac{1}{sr}  )(1 - \tfrac{s}{ r})} \tfrac{1}{sr} \tfrac{s}{ r}
\end{align}
with an extra fugacity $t$ that keeps track of the picture number. It is not hard to see that the four generating functions are actually identical suggesting that they count the same dimensions in different representations. The total partition function can thus be simplified to 
\begin{equation}
    \mathbb{P}_{\Omega^{(\bullet|\bullet)}} =
    (1-t)^2 
        \mathbb{P}_{\Omega^{(\bullet|0)}} \, .
\end{equation}
In the next subsections we analyze different sectors separately.

\subsection{Picture two and Hodge duality}
Having studied the picture zero sector in the previous section we now perform an analogous analysis for picture two states. Furthermore we construct an explicit map between said two sectors and argue it being a quasi-isomorphism.
The states in the picture two sector are generated freely as
\begin{equation}
    c^{p_2}\psi^{\tiny{\hash}\psi} \delta^{(m)}(\gamma)\delta^{(n)}(\beta)
\end{equation}
with $R$-charge  $\tiny{\hash}\psi-m-n-2$. The partition function enumerating these states is given by 
\begin{eqnarray}
\label{ppC}
\mathbb{P}_{\Omega^{(\bullet|2)}} (r,s) = \frac{(1 +  r)^D(1 + s)}{(1 - \tfrac{1}{sr}  )(1 -\tfrac{s}{ r})}
 \frac{1}{sr}  \frac{s}{ r}\,,
 \end{eqnarray}
where the last factors $\frac{1}{sr}  \frac{s}{ r}$ account for the gradings  of $\delta(\gamma)\delta(\beta)$. Upon subsuming these factors in the denominator one sees that
\begin{eqnarray}
\label{ppD}
\mathbb{P}_{\Omega^{(\bullet|2)}}(r,s) = \mathbb{P}_{\Omega^{(\bullet|0)}}(r,s) \,,
\end{eqnarray}
which shows that the states counted in picture zero and picture two are in one to one correspondence. 
The same is true for the cohomology of the BRST operator $Q$, except that the $R$-charge is mapped to $-R-2$. 
Another way to justify the relation between picture zero and picture two sectors is through the action of a Hodge operator $\star_{BV}$, defined as follows 
\begin{gather}\label{eq:star_BV}
\star_{BV} \psi \star_{BV} = \partial_{\psi} 
\quad
\star_{BV} \beta \star_{BV} = \partial_{\gamma}
\quad
\star_{BV} \gamma \star_{BV} = \partial_{\beta} \, ,
\end{gather}
which is an extension of a regular $\star$ action only on the fields $\psi$. Applied on states from the picture zero sector
\begin{equation}
    c^{p_2}\psi^q \delta^{(m)}(\gamma)\delta^{(n)}(\beta)
\end{equation}
one gets picture two states:
\begin{equation}
    c^{p_2} \gamma^n \beta^m ( \psi)^{d-q} \, .
\end{equation} 
Since furthermore $Q$ commutes with $\star_{BV}$ the map ascends to a map on cohomologies, where it actually provides an isomorphism of cohomologies.

\subsection{Picture one}
While the $R$-charge is bounded either from below or above in picture $0$ and picture $2$, picture one is unbounded in $R$-charge. Nevertheless the respective cohomologies are isomorphic for $R\geq 0$ and counted by the refined Euler characteristics as suggested by \eqref{eq:fullp}.  Qualitatively, however, the off-shell spectrum is different for negative and positive R charge parts of the spectrum as we will show concretely. In picture 1, more precisely in $\Omega^{(\bullet|1)}_\beta$ we have 
\begin{align}\label{eq:pf_p10}
\ket{\phi}
= & \cdots\nonumber\\
&+\delta^{(2)}(\beta)\epsilon_0^{ (R+3)}\cdot\psi^{(R+3)}+\delta^{(1)}(\beta)\gamma\epsilon_0^{ (R+1)}\cdot\psi^{(R+1)}+\delta(\beta)\gamma^2\epsilon_0^{ (R)}\cdot\psi^{R}\nonumber \\
&+\delta^{(1)}(\beta)\epsilon_0^{ (R+2)}\cdot\psi^{(R+2)}+\delta(\beta)\gamma\epsilon_0^{ (R)}\cdot\psi^{R}\nonumber \\
&+\delta(\beta)\epsilon_0^{ (R+1)}\cdot\psi^{(R+1)}\,,
\end{align}
where we ordered the states in decreasing ghost degree $k$. The presence of $\delta(\beta)$ implies that $Q$-closure requires the polarization tensors to be closed under the exterior differential which means that it has to be $d$-exact. This is true also for the last line except that there is no such state of ghost number zero in the complex, i.e. $B_1=\emptyset$. Thus $\mathbb{H}_1=Z_1$. Furthermore,  $\mathbb{H}_k=\emptyset$ for $k>1$. We have thus shown that the for R-charge $R\geq 0$, the cohomologies in picture $0$ and picture $(1,0)$ are isomorphic. The map between them is given by the picture changing operator 
\begin{align}
Y(\beta)=\left[Q,\Theta(\beta)\right]=\delta(\beta)\psi\cdot k+\delta(\beta)b\gamma\,.
\label{PCO:increase-pic}
\end{align}
The second term acts trivially on $\mathbb{H}_0$ in picture $0$ while the first maps the potential in $\mathbb{H}_0$ to its field strength in $\mathbb{H}_1$ picture $1$. We should note though that as complexes the different pictures are not isomorphic. In particular, the states in \eqref{eq:pf_p10} with differentiated delta functions are not in the image of $Y(\beta)$.

Picture $\Omega^{(\bullet|1)}_\gamma$ is in turn obtained from picture $2$ by picture changing. A generic state is
\begin{align}\label{eq:pf_p01}
\ket{\phi}
=&\delta(\gamma)\epsilon_0^{ (R+1)}\cdot\psi^{(R+1)}\nonumber\\
&+\delta^{(1)}(\gamma)\epsilon_0^{ (R+2)}\cdot\psi^{(R+2)}+\delta(\gamma)\beta\epsilon_0^{(r-1)}\cdot\psi^{ (R-1)}\nonumber \\
&+\cdots
\end{align}
with the isomorphism 
given by the inverse picture changing operator 
\begin{align}
Z(\beta)= -\delta^\prime(\tfrac{\partial }{\partial \beta}) \tfrac{\partial }{\partial \gamma}\bar\psi\cdot k+\delta(\tfrac{\partial }{\partial \beta})\tfrac{\partial }{\partial c} \tfrac{\partial }{\partial \gamma}\,.
\end{align}
The full partition function in the picture one sector then reads
 \begin{gather}
   \mathbb{P}(\Omega_{\gamma}^{(\bullet|1)})
   +
    \mathbb{P}(\Omega_{\beta}^{(\bullet|1)})
= \frac{(1+r)^D(1+s)}{(1-\frac{1}{sr})(1-\frac{r}{s})} \frac{1}{sr} + \frac{(1+r)^D(1+s)}{(1-sr)(1-\frac{s}{r})} \frac{s}{r} \, . 
\end{gather}

\subsubsection{Negative $R$ charge}
We now show that the negative $R$ charge part of the picture one sector contains a topological sector with a Hilbert space  equivalent to Chern-Simons theory.
Let us fix $R=-(K+1)$ and dimension $D$. The list of the states is 
\begin{equation}
    \begin{array}{cccc}
        \delta^{K}(\gamma) &  & \\
        \beta \delta^{K+1}(\gamma) & \delta^{K+1}(\gamma) \psi^m & \\
         \beta^2 \delta^{K+2}(\gamma) & \beta \delta^{K+2}(\gamma) \psi^m  & \delta^{K+2}(\gamma) \psi^{m_1} \psi^{m_2} 
         \\
         \vdots & \vdots & &
         \\
         \beta^D \delta^{K+D}(\gamma) & \beta^{D-1} \delta^{K+2}(\gamma) \psi^m  &  \ldots & \delta^{K+D}(\gamma) \psi^{D}
         \\
                  \vdots & \vdots & & \vdots
         \\
         \beta^{D+1} \delta^{K+D+1}(\gamma) & \beta^D \delta^{K+2}(\gamma) \psi^m  & \ldots &   \beta \delta^{K+D}(\gamma) \psi^{D} \, .
         \\
    \end{array}
\end{equation}
Upon introducing two operators
\begin{equation}
    \psi \partial_{\gamma} \qquad
    \beta \partial_{\gamma}
\end{equation}
we see that the spectrum is spanned freely: The first operator shifts along the diagonals whereas the second one goes vertically downwards and spans the entire spectrum.
The operators are respectively fermionic and bosonic.

Now it is sufficient to count them with their appropriate degrees. Recalling Table \ref{redefined q.n.}, $\psi \partial_\gamma$ has charge $\frac{1}{s}$, while $\beta \partial_\gamma$ has charge $\frac{1}{s^2}$. The partition function therefore reads
\begin{equation}
 \mathbb{P}(s)= \frac{(1+ \frac{1}{s})^D}{1 - \frac{1}{s^2}} \tfrac{1}{sr} \, .
\end{equation}

Thus we can see that the numerator (responsible of enumerating the states on diagonals) resembles a copy of a Chern--Simons-like system. This Chern--Simons subsector has an infinite multiplicity (one for each diagonal).
Furthermore on the first diagonal (starting from $\delta^{K+D}(\gamma)\psi^D$ and and going back up to $\delta^K(\gamma)$), $Q$ acts as $Q_{CS}$ since the states are independent of  $\beta$.

\subsection{Reduction to $\mathcal{N}=1$}
There is a canonical embedding of the $\mathcal{N}=1$ world line into the $\mathcal{N}=2$ model by taking the real part of the supercharges. The $\mathcal{N}=1$ case has been studied in several works (see for example  \cite{Howe:1988ft,Brink:1976uf, Casalbuoni:1974pj,Casalbuoni:1975hx} and the references therein), in the present section we describe along the previous analysis how the $\mathcal{N}=1$ case is embedded into the 
$\mathcal{N}=2$ case. We leave to a forthcoming paper a new analysis on $\mathcal{N}=1$ spinning particle tackling the problem of pictures in that case \cite{newIVOgroup}.

At ghost number zero we have a  gauge-equivalent representation of $\mathbb{H}_d(  V_{p_1} )$
in ${\Omega^{(\bullet|0)}}$ 
given by
\begin{align}\label{eq:cog=ge}
\{(\epsilon^{(n)}+k\wedge\lambda^{(n-1)})\cdot\psi^{ n}+\gamma\beta (k\cdot \lambda^{(n+1)})\cdot\psi^{n}\}
\end{align}
as one can deduce from \eqref{eq:cog}.
We choose the gauge parameter $\sum\limits_{p}\beta \lambda^{(p)}\cdot\psi^{p}$ with $\lambda$ on-shell,  such that 
\begin{align}
    \epsilon^{(n)}=k\cdot \lambda^{(n+1)}\,, \quad n=0,\cdots,n-1\,.
\end{align}
Then \eqref{eq:cog=ge} becomes 
\begin{align}\label{eq:cog=ge1}
\{(\epsilon^{(n)}+k\wedge\lambda^{(n-1)})\cdot\psi^{ n}+\gamma\beta \epsilon^{(n)}\cdot\psi^{ n}\}\,.
\end{align}
We then repeat this with gauge parameter $\sum\limits_{p}\frac{(\beta\gamma)}{2}\beta \lambda^{(p)}\cdot\psi^{p}$, giving 
\begin{align}\label{eq:cog=ge1}
\{(\epsilon^{(n)}+k\wedge\lambda^{(n-1)})\cdot\psi^{ n}+\gamma\beta (\epsilon^{(n)}+k\wedge\lambda^{(n-1)})\cdot\psi^{n}+\frac{(\beta\gamma)^2}{2} \epsilon^{(n)}\cdot\psi^{n}\}
\end{align}
and so on, summing up to
\begin{align}\label{eq:sum}
    e^{\beta\gamma} \varepsilon^{(n)}\cdot\psi^{n}\quad\text{with}\quad k\cdot \varepsilon^{n}+k\wedge \varepsilon^{(n-2)}=0\,.
\end{align}
If we stop this gauge transformation at finite order in $(\beta\gamma)$, then in the highest power, $\varepsilon^{(n)}$ is replaced by $\epsilon^{(n)}$. On the other hand, we have 
\begin{align}
    Q e^{\beta\gamma} \varepsilon^{(n)}\cdot\psi^{ n}= e^{\beta\gamma}\left(\gamma(q+\bar q)+b\gamma^2 \right) \varepsilon^{(n)}\cdot\psi^{ n}\,,
\end{align}
where the expression in the bracket is recognized as the on-shell BRST operator $Q_{\mathcal{N}=1}$ of the $\mathcal{N}=1$ spinning world line. Conversely, the constraint \eqref{eq:sum} characterizes generic elements in the kernel of $Q_{\mathcal{N}=1}$. Furthermore, if 
\begin{align}
    \varepsilon(\psi)= \sum\limits_n\varepsilon^{(n)}\cdot\psi^{ n}
\end{align}
is $Q$-exact, $ \varepsilon(\psi)=Q_{\mathcal{N}=1}\lambda(\psi)$, then $\epsilon(\psi)=e^{\beta\gamma}\varepsilon(\psi)$ is $Q$-exact. Thus, the gauge transformation (\ref{eq:cog=ge}-\ref{eq:sum}) maps elements in cohomology of $Q_{\mathcal{N}=2}$ to elements in cohomology of $Q_{\mathcal{N}=1}$. Note that the map \eqref{eq:sum} does not preserve the $R$-charge, as expected since there is no $R$ symmetry for $\mathcal{N}=1$. Consequently, when computing the partition function we should sum over all values of $p_1$, by setting $r=1$. Then \eqref{ppA} becomes
 \begin{eqnarray}
\label{ppAr}
\partial_s \mathbb{P}_{\Omega^{(\bullet|0)}} (1,s)|_{s=-1} = 2^{D-2}  
\end{eqnarray}
and counts the dimension of  $\mathbb{H}_{Q_{\mathcal{N}=1}}(V)$ at ghost degree $0$. For instance, in 4-dimensions, this decomposes into a scalar $(1)$, a transverse vector $(2)$, a transverse two-form $(1)$ and a transverse 3-form $(0)$. 

One can similarly identify a subspace that describes the cohomology of the $\mathcal{N}=0$ particle by restricting to $p_3=0$. In that case only the first term in the expansion of \eqref{cippaBBB} representing the massless scalar particle is described by that model.

\subsection{$\mathcal{N}>2$}  \label{ext}
It is not hard to convince oneself that  the derivation in section \ref{sec:p0} goes through with small modifications for any $\mathcal{N}\in 2\mathbb{Z}$. We can thus apply \eqref{eq:partQ} or \eqref{ppA} for any $\mathcal{N}>2$ with the difference that in addition to $R$ there will be additional conserved quantum numbers for each generator of the maximal torus of the $R$- symmetry group. For instance, for $\mathcal{N}=4$ we have 
\begin{align}\label{cippaBBB2}
  \mathbb{P}_ V (
  r_1,r_2,s)  
={\bf tr}_{ V }[s^{\mathbf{k}}r_1^{\mathbf{R}_1}r_2^{\mathbf{R}_2}
  ]
&=\frac{(1+r_1)^D(1+r_2)^D(1+ s)}{(1- sr_1)(1-\frac{r_1}{ s})(1- sr_2)(1-\frac{r_2}{ s})} \, .
\end{align}
where $R_1,R_2$ measure the charges with respect to the $U(1)\times U(1)$ maximal torus of $U(2)$. As an example, we consider $R_1=R_2=1$. Then 
\begin{align}
    r_1 r_2 \left(D^2+2 D s+\frac{2 D}{s}+s^2+\frac{1}{s^2}+2\right)
\end{align}
evaluated at $s=-1$ counts the degrees of freedom of the graviton, Kalb-Ramond field and the dilaton in $D$-dimension. Furthermore, one can read off the gauge and gauge for gauge degrees of freedom from the inverse powers in $s$. The detailed dynamics of this system was described in \cite{Bonezzi:2018box}. 

\section{Another Look at the BRST Cohomology: Off-Shell Construction}
\label{sec:BRST-coh}
Having presented a general derivation of the spectrum from the partition function point of view when the Hamiltonian constraint $P^2 =0$ is imposed, here we give an alternative perspective on the BRST cohomology computation which is useful for extracting the explicit equations of motion.

To do this systematically we observe that the 
BRST differential can be decomposed into the following pieces
\begin{eqnarray}
    \label{AA1}
    Q = Q_2 + Q_1 + Q_0 = 
    c P^2 + \gamma \bar\psi \cdot P + \psi\cdot P \frac{\partial}{\partial \beta} - \gamma  \frac{\partial}{\partial \beta}  \frac{\partial}{\partial c} \, ,
\end{eqnarray}
graded by a charge acting on the ghost fields as follows
\begin{eqnarray}
    \label{AA2}
    \gamma \rightarrow  s \gamma\,, ~~~~~
    \beta \rightarrow s^{-1} \beta \,, ~~~~
    c \rightarrow s^2 c\,. 
\end{eqnarray}
The splitting of the nilpotency conditions $Q^2=0$ imply the following condition of each separated piece 
\begin{equation}
    Q_0^2=0\,, ~~~~~~~ \{Q_0, Q_1\} = 0\,, ~~~~~~~Q_1^2 + \{Q_2,Q_0\} =0\,, ~~~~~
    \{Q_1, Q_2\} =0\,, ~~~~ Q_2^2 =0
\end{equation}
from which it is clear that $Q_0$ and $Q_2$ are two nilpotent operators. This allows for the following strategy (note that the strategy is alternative to the one presented in the previous sections): firstly, analyze the cohomology of $Q_0$, $\mathbb{H}_{Q_0}(V)$ on the full Fock space $V$. Secondly, study the cohomology of the full BRST operator as a relative cohomology with respect to $Q_0$. Indeed, $Q_1$ commutes with $Q_0$ and $(Q_1)^2 \sim 0$ on the cohomology of $Q_0$. This permits the construction of the relative cohomology $\mathbb{H}_{Q_1}( \mathbb{H}_{Q_0} (V))$. Finally, we can compute the cohomology of $Q_2$ on $\mathbb{H}_{Q_1}( \mathbb{H}_{Q_0} (V))$ to obtain the full cohomology. In the next section, we first compute the cohomology of $Q_0$ and 
then we show how to retrieve the full cohomology by computing the differential equations for the component fields. 

\subsection{$Q_0$ cohomology}
Here we analyze the $Q_0$ cohomology which involves only the ghost variables. The rest of the variables play a spectator role here. The Hilbert space is defined by a generic state (in any picture) expanded into a linear combination of 
\begin{eqnarray}
    \label{AA4}
    \gamma^p, \beta^p, \delta^{(p)}(\gamma), \delta^{(p)}(\beta)  
\end{eqnarray}
with coefficients in $x,\psi$. 
Since $\delta(\gamma)$ and $\delta(\beta)$ are anticommuting quantities, they can only appear once or multiplied as $\delta(\gamma)\delta(\beta)$. Notice that for $\delta(\gamma)$ and $\delta(\beta)$ distribution properties such as $\gamma \delta'(\gamma) = - \delta(\gamma)$ are valid and taken into account in the derivation. 

We separate the cohomology in three sectors: picture zero, picture one and 
picture two which are constructed as follows 
\begin{eqnarray}
    \label{AA5}
    \omega_{\text{I}} &=& \sum_{p,q} \left( A_{p,q} \,\beta^p \gamma^q + 
    \Xi_{p,q} \, c \,\beta^p \gamma^q \right) \\
    \omega_{\text{II/III}} &=& \sum_{p,q} \left( B_{p,q} \,\beta^p \,\delta^{(q)}(\gamma) + 
    \Gamma_{p,q} \, c \,\beta^p \delta^{(q)}(\gamma) \right) 
    + 
    \sum_{p,q} \left( C_{p,q} \,\delta^{(p)}(\beta)\gamma^q + 
    \Sigma_{p,q} \, c \, \delta^{(p)}(\beta) \gamma^q \right)
    \\
    \omega_{\text{IV}} &=& \sum_{p,q} \left( \Delta_{p,q} \,\delta^{(p)}(\beta) \delta^{(q)}(\gamma) + 
    D_{p,q} \, c \,\delta^{(p)}(\beta) \delta^{(q)}(\gamma) \right) 
    \end{eqnarray}
with $p,q \geq 0$. 
The fields $A_{p,q}, \dots, \Delta_{p,q}$ are depending upon $x^\mu, \psi^\mu$. In the following we neglect this dependence and we consider only the ghost degree of freedom. 

By brute force computation of the cohomology we found the following expressions
\begin{eqnarray}
    \label{AA6}
    \omega_{\text{I}} &=& \sum_{p\geq 0} A_{p,0} \beta^p + 
    \sum_{q\geq 0} \chi_{0,q} \, c \gamma^q \\
    \label{AA7}
    \omega_{\text{II}} &=& 
     C_{0,0} \delta(\beta) + 
     \sum_{p>0} C_{p,0} \delta^{(p)}(\beta) + 
     \sum_{q>0} C_{0,q} \delta(\beta) \gamma^q  \\
    \label{AA8}
    \omega_{\text{III}} &=& 
     c \left(\Gamma_{0,0} \delta(\gamma) + 
     \sum_{p>0}  \Gamma_{p,0} \delta^{(p)}(\gamma) + 
     \sum_{q>0} \Gamma_{0,q} \delta(\gamma) \beta^q \right)  \\
  \label{AA9}
     \omega_{\text{IV}} &=& \sum_{p\geq 0} D_{p,0} \,c\, \delta^{(p)}(\beta) \delta(\gamma) + 
    \sum_{q\geq 0} \Delta_{0,q}  \,\delta(\beta) \delta^{(q)}(\gamma)
\end{eqnarray}

Then, the cohomology can be enumerated by a twisted partition function, where the twisting introduces an extra minus sign on the ghosts, compared to previous formulas: 
\begin{eqnarray}
    \label{AB1}
    {\rm T} \mathbb{P}(r,s) = \frac{(1-s^2)}{(1- r s)(1 - r /s)}\,.
\end{eqnarray}
Here we kept the extra fugacity for $\beta$ and $\gamma$ for later convenience. Let us check whether this can be achieved also by counting the cohomologies in \eqref{AA6}, \eqref{AA7}, \eqref{AA8}, and \eqref{AA9}. 

The picture zero $\omega_{\text{I}}$: 
\begin{eqnarray}
    \label{AB2}
     {\rm T} \mathbb{P}_{\text I}(r,s) = \frac{1}{1- r/s} - s^2 \frac{1}{1- r s} =  {\rm T} \mathbb{P}(r,s) 
\end{eqnarray}
where the factor $(-s^2)$ in the second term comes from the anticommuting ghost $c$ appearing in 
\eqref{AA6}. 

The picture one $\omega_{\text{II}}$:
\begin{eqnarray}
    \label{AB3}
    {\rm T}  \mathbb{P}_{\text{II}}(r,s) = \left(- \frac{s}{r}\right)
     \left(
     1 + \frac{s/r}{1- s/r} + \frac{r s}{1- r s} \right)  =  {\rm T} \mathbb{P}(r,s) 
\end{eqnarray}
where the factor $\left(- \frac{s}{r}\right)$ takes into account $\delta(\beta)$ 
contribution, while $\frac{s/r}{1- s/r}$ denotes the power of derivatives on  $\delta(\beta)$. 

The picture one $\omega_{\text{III}}$:
\begin{eqnarray}
    \label{AB4}
    {\rm T}  \mathbb{P}_{\text{III}}(r,s) = (-s^2) \left(- \frac{1}{s r}\right)
     \left(
     1 + \frac{1/s r}{1- 1/s r} + \frac{r/s}{1- r/s} \right)  = {\rm T} \mathbb{P}(r,s) 
\end{eqnarray}
where the factor $\left(- \frac{1}{s r}\right)$ takes into account $\delta(\gamma)$ 
contribution, while $\frac{1/s r}{1- 1/s r}$ denotes the power of derivatives on  $\delta(\gamma)$.  Finally, the factor $(-s^2)$ takes into account the ghost $c$ in front. 

The picture two $\omega_{\text{IV}}$:
\begin{eqnarray}
    \label{AB5}
    {\rm T}  \mathbb{P}_{\text{IV}}(r,s) = \left(- \frac{1}{s r}\right)\left(- \frac{s}{r}\right)
     \left(
     \frac{1}{1- 1/s r} - s^2 \frac{1}{1- s/r}\right)  =  {\rm T}\mathbb{P}(r,s) 
\end{eqnarray}
where the factor $\left(- \frac{1}{s r}\right)\left(- \frac{s}{r}\right)$ takes into account $\delta(\gamma) \delta(\gamma)$ 
contribution and the factor $(-s^2)$ takes into account the ghost $c$. Thus the computation confirms that the cohomology is identical in all sectors even though the off shell states are different. However there is a crucial aspect to comment on. Comparing with the partition function in the previous section, we see that the factor due to the ghost $c$ is linear in $s$ and not quadratic, preventing the natural identification choice which would instead work for the denominators. Such mismatch will be elucidated in the forthcoming section.

Note that we have split the picture one  sector into two parts generated by $\delta(\gamma)$ and $\delta(\beta)$ respectively. However this splitting is not preserved by general covariance of the model. Furthermore notice that for $\omega_{\text{I}}$ and $\omega_{\text{IV}}$ there are two sectors of contributions, namely even and odd cohomologies, whereas in the picture one there is only one type of parity. Therefore, even though the counting seems to be the same the parity is not respected.  

In order to repair this problem, we observe that there should be a mapping between the different spaces and since they are isomorphic there should be an isomorphism to connect those spaces. Such an application is called {\it Picture Changing Operator} PCO. PCOs are closed and not-exact operators\footnote{Lack of exactness is related to the space under inspection. Indeed in the following examples the PCOs will appear as $Q_0$-exact, but only upon introducing the Heaviside theta.}. For the case at hand, there are indeed four of them, more precisely
\begin{eqnarray}
    \label{AB6}
    \hspace{1.5cm}
    Y_0 (\gamma) = c\delta(\gamma)\,, \hspace{2.0cm}
    Y_0 (\beta) = \left[Q_0, \frac{\Theta(\beta)}{\gamma}\right] =  \delta(\beta) \frac{\partial}{\partial c}
\end{eqnarray}
which raise the picture and 
\begin{eqnarray}
    \label{AB7}
    Z_0 (\gamma) = \left[Q_0, \beta\Theta\left(\frac{\partial}{\partial \gamma}\right)\right] = \delta
    \left(\frac{\partial}{\partial \gamma}\right) \frac{\partial}{\partial c} \,, \hspace{1.0cm}
    Z_0 (\beta) = c \, \delta
    \left(\frac{\partial}{\partial \beta}\right)
\end{eqnarray}
which lower the picture and  where we used the identity $\gamma \Theta(\partial/\partial \gamma) = \delta(\partial/\partial \gamma)$. Note the the PCO $Y(\gamma)$ and $Y(\beta)$ lower the $r$-charge, while the PCO $Z(\gamma)$ and $Z(\beta)$ increase the $r$-charge by one unit. Notice that the choice is not unique as there exists other possible PCOs, for instance 
\begin{eqnarray}
    \label{AB67}
    \hspace{1.5cm}
    Y_0 (\gamma) = c\beta\delta(\gamma)\,, \hspace{2.0cm}
    Y_0 (\beta) = \left[Q_0, \Theta(\beta)\right] =  \gamma\delta(\beta) \frac{\partial}{\partial c} 
\end{eqnarray}
which raise the picture and 
\begin{eqnarray}
    \label{AB7}
    Z_0 (\gamma) = \left[Q_0, \Theta\left(\frac{\partial}{\partial \gamma}\right)\right] = \delta
    \left(\frac{\partial}{\partial \gamma}\right) \frac{\partial}{\partial c}  \frac{\partial}{\partial \beta}\,, \hspace{1.0cm}
    Z_0(\beta) = c \, \delta
    \left(\frac{\partial}{\partial \beta}\right)\frac{\partial}{\partial \gamma}
\end{eqnarray} that lowers the picture. This quartet of operators does not change the $r$-charge. These PCO will turn out to be useful when we add the matter to the spectrum. 
 
In addition, they satisfy 
\begin{eqnarray}
    Z_0 (\gamma) \,Y_0 (\gamma) Z_0 (\gamma) =Z_0 (\gamma)
\,, \hspace{1.5cm}  Y_0 (\gamma) Z_0 (\gamma) \, Y_0 (\gamma) = Y_0 (\gamma) \nonumber \\
  Z_0 (\beta) \,Y_0 (\beta) Z_0 (\beta) =Z_0(\beta)
\,, \hspace{1.5cm}  Y_0 (\beta) Z_0 (\beta) \, Y_0 (\beta) = Y_0 (\beta)
\end{eqnarray}
which is a generalized inverse formula. In several applications, that generalized formula is sufficient for several derivations. Since the PCOs are cohomological classes they differ by exact terms from each other, and therefore one can find the exact pieces without kernel for which there is an inverse. 

Notice that two of them, $Y_0(\beta)$ and $Z_0(\gamma)$, are written as BRST exact of the Heaviside theta function $\Theta$. The second set depends upon the differential operators $\delta
    \left(\frac{\partial}{\partial \gamma}\right), \delta\left(\frac{\partial}{\partial \beta}\right)$ which 
    can be conveniently defined using the integral representation of the Dirac delta 
\begin{eqnarray}
\label{AB8}
\delta\left(\frac{\partial}{\partial \beta}\right) = 
\int_{-\infty}^\infty du e^{i u \frac{\partial}{\partial \beta}}
\end{eqnarray}
leading to 
\begin{eqnarray}
    \label{AA9}
    \delta\left(\frac{\partial}{\partial \beta}\right) \delta(\beta) = 1\,, 
    \hspace{1.5cm}
    \delta\left(\frac{\partial}{\partial \gamma}\right) \delta(\gamma) = 1\,, 
\end{eqnarray}

Then, we get 
\begin{eqnarray}\label{AA10}
    Y_0(\gamma) \omega_{\text{I}} &=&\sum_{p\geq 0} A_{p,0} \, \beta^p   c \delta(\gamma)  \subset \omega_{\text{III}}\nonumber \\
    Y_0(\beta)  \omega_{\text{I}} &=& 
     \sum_{q\geq 0} \chi_{0,q} \, \gamma^{(q)} \delta(
    \beta) \subset \omega_{\text{II}} \nonumber \\
    Z_0(\gamma) \omega_{\text{IV}} &=&  
     \sum_{p\geq 0} D_{p,0} \, \delta^{(p)}(\beta) \subset  \omega_{\text{II}} \nonumber \\
     Z_0(\beta) \omega_{\text{IV}} &=& 
      \sum_{q\geq 0} \Delta_{0,q}   \, c \delta^{(q)}(\gamma) 
       \subset  \omega_{\text{III}}     
\end{eqnarray}
where we have indicated that the pieces obtained by acting with a PCO on $\omega_{\text{I}}$ and $\omega_{\text{IV}}$ are not complete but a subset of the picture-one contributions. However, putting together this observation and the remark done above we see that it is convenient to redefine the picture-one cohomologies as follows
\begin{eqnarray}
    \label{AA11}
     \omega'_{\text{II}} &=& 
       \sum_{q\geq 0} C_{0,q}\,  \delta(\beta) \gamma^q +
            \sum_{q\geq 0} \Gamma_{0,q} \, c \delta(\gamma) \beta^q
      \\
    \label{AA12}
    \omega'_{\text{III}} &=& 
          \sum_{p\geq 0} C_{p,0} \, \delta^{(p)}(\beta) +
      \sum_{p \geq 0}  \Gamma_{p,0} \,  c \delta^{(p)}(\gamma)
\end{eqnarray}
where we mixed the picture with $\delta(\beta)$ with those with $\delta(\gamma)$. In this way, we see that 
\begin{eqnarray}
    \label{AA13}
    (Y_0 (\gamma) + Y_0 (\beta)) 
    \omega_{\text{I}} = \omega'_{\text{III}}\,, \nonumber \\
    (Z_0 (\gamma) + Z_0 (\beta)) 
    \omega_{\text{IV}} = \omega'_{\text{II}}\,, \nonumber 
\end{eqnarray}
mapping isomorphically $\omega_{\text{I}}$ into
$\omega'_{\text{III}}$ and $\omega_{\text{IV}}$ into 
$\omega_{\text{II}}'$. 
Finally in order to map the cohomology $\omega_{\text{I}}$ into the cohomology $\omega_{\text{IV}}$ we use the Hodge duality 
\begin{eqnarray}
    \label{AA14}
    &&\star_{BV_0} \omega_{\text{I}}(c,\gamma,\beta) = 
    i^\#\int dx dy e^{i ( x \gamma + y \beta) } \omega_{\text{I}}(c, x,y)   \subset \omega_{\text{IV}} \nonumber \\
    &&\star_{BV_0} \omega_{\text{IV}} \subset \omega_{\text{I}}
\end{eqnarray}
where the exponent $\#$ is fixed by requiring the idem-potency of $\star_{BV}$ dual operation. 

Therefore, these properties show that all spaces are isomorphic and we explicit construct the mapping. Notice that the 
rearrangements of the picture-one cohomologies takes care also of the problem of parity. However, in order to 
verify that this rearrangement reproduces the correct partition function we use the same computation as above to get 
\begin{eqnarray}
    \label{AA15}
    {\rm T}\mathbb{P}'_{\text{II}}(r,s) &=& 
    - s^2 \left(-\frac{1}{r s} \right)
        \frac{1}{1- r/s}  + 
        \left(-\frac{s}{r} \right)
        \frac{1}{1- r s} = {\rm T}\mathbb{P}(r,s)
\nonumber \\
 {\rm T} \mathbb{P}'_{\text{III}}(r,s) &=& 
    - s^2 \left(-\frac{1}{r s} \right)
        \frac{1}{1- 1/r s}  + 
        \left(-\frac{s}{r} \right)
        \frac{1}{1- s/r} = {\rm T}\mathbb{P}(r,s)
\end{eqnarray}
which prove the equivalence. 

\subsection{Back to full cohomology} 
\label{Back2fullcoho}
Having investigated the $Q_0$-cohomology we return to the question of full $Q$-cohomology. We only give an explicit analysis of the zero picture sector as the analysis proceeds similarly in other parts of the spectrum. 

Since the $\psi$'s carry degree $1$ of $r$-charge, a fixed $r$-charge state in picture zero sector reads
\begin{eqnarray}
    \label{matA}
    \omega_{\text{I}}^r = \sum_{p,q;\, p+q=0}^{p+q=r} 
    A^{r-p-q}_{p,q}(\psi,x) \beta^p \gamma^q + \chi^{r-p-q}_{p,q}(\psi,x) c 
\beta^p \gamma^q
\end{eqnarray}
where $A^r_{p,q}(\psi,x), \chi^r_{p,q}(\psi,x)$ are 
$r$ forms (e.g.~$A^r_{p,q}(\psi,x) = A_{p,q \, \mu_1 \cdots \mu_r}(x) \psi^{\mu_1} \cdots \psi^{\mu_r}$). 
BRST closure yields the following equations on the states in the $r=0$ sector.
\begin{eqnarray}
    \label{matB}
    \omega_{\text{I}}^{r=0} = 
    A^0_{0,0}(x) + \chi^0_{0,0}(x) c 
\end{eqnarray}
where $A^0_{0,0}(x)$ and $\chi^0_{0,0}(x)$ are $0$-forms. 
They satisfy ($\square =\partial^2$)
\begin{eqnarray}
    \label{matC}
    \partial^2  A^0_{0,0} =0\,, 
\end{eqnarray}
from the BRST closure and 
\begin{eqnarray}
    \label{matD}
    \delta A^0_{0,0} =0\,, \hspace{1.5cm}
    \delta \chi^0_{0,0} = \partial^2 \Sigma^0_{0,0}
\end{eqnarray}
where $\Sigma^0_{0,0}$ is a gauge parameter. This shows that we can remove the fields $\chi^0_{0,0}$ using the gauge parameter $\Sigma^0_{0,0}$ only if $\partial^2 \chi^0_{0,0} \neq 0$, otherwise, if $\partial^2 \chi^0_{0,0} =0$ it cannot be removed by the gauge transformation and therefore there are two degrees of freedom with Klein-Gordon equations. 

Let us consider the case $r=1$: 
\begin{eqnarray}
    \label{matE}
    \omega_{\text{I}}^{r=1} = 
    \Big(A^1_{0,0}({\psi}, x) + A^{0}_{1,0}(x) \beta + A^0_{0,1}(x) \gamma\Big) 
    +\Big(\chi^1_{0,0}({\psi},x) + \chi^{0}_{1,0}(x) \beta + 
    \chi^0_{0,1}(x) \gamma\Big) c
\end{eqnarray}
 leading to the equations
 \begin{eqnarray}
     \label{matE}
     && d^\dagger A^1_{0,0} + \chi^0_{1,0} =0 \nonumber \\
     && \partial^2 A^1_{0,0} - d \chi^0_{1,0} =0 \nonumber \\
     && \partial^2 A^0_{0,1} - d^\dagger 
     \chi^1_{0,0} =0 \nonumber \\
     && d A^0_{1,0} =0 \nonumber \\
     && \partial^2  A^0_{1,0} =0      
 \end{eqnarray}
and, 
the gauge symmetry
\begin{eqnarray}
    \label{matF}
    &&\delta A^1_{0,0} = d \Sigma^0_{1,0} \nonumber \\
    &&\delta A^0_{1,0} =0 \nonumber \\
    && \delta A^0_{0,1} = d^\dagger \Sigma^1_{0,0} + M^0_{1,0} \nonumber \\
    && \delta \chi^1_{0,0} = \partial^2 \Sigma^1_{0,0} - d M^0_{0,1} \nonumber \\
    && \delta \chi^0_{1,0} = \partial^2 \Sigma^0_{1,0} \nonumber \\
    && \delta \chi^0_{0,1} = - d^\dagger M^1_{0,0} + \partial^2 \Sigma^0_{0,1}
\end{eqnarray}
Inserting the first line of \eqref{matE} into the 
second line and using { $\partial^2 = -(d d^\dagger + d^\dagger d)$}, we get
\begin{eqnarray}
    \label{matG}
    - d^\dagger d A^1_{0,0} =0\,, 
\end{eqnarray}
which is Maxwell equation. The gauge symmetry described in \eqref{matF} indeed is the usual abelian gauge symmetry for a
$1$-form field. The fourth equation tells us that $A^0_{1,0}$ is constant and finally we can recast the third line as follows 
\begin{eqnarray}
\label{matH}
- d^\dagger (dA^0_{0,1} + \chi^1_{0,0}) =
- d^\dagger \hat \chi^1_{0,0} =0
\end{eqnarray}
where $\hat \chi^1_{0,0} =  \chi^1_{0,0} + 
dA^0_{0,1}$ where the $d$-part is reabsorbed by this redefinition. The gauge transformation of the redefined field is 
\begin{eqnarray}
    \label{matI}
 \delta \hat\chi^1_{0,0} = - d^\dagger d \, \Sigma^1_{0,0}     
\end{eqnarray}
which implies that the field $\hat\chi^1_{0,0}$ can be gauged away if $\hat\chi^1_{0,0}$ is not in the kernel of $d^\dagger d$. Vice versa, if  $\hat\chi^1_{0,0}$ is in the kernel of $d^\dagger d$, it cannot be gauged away and it describes a second copy of Maxwell $1$-form. Notice this matches with the partition function \eqref{ppA} 
at $R$-symmetry with $r=1$: 
\begin{eqnarray}
    \label{matL}
   {\rm T} \mathbb{P}_{r=1}( r, s) = 
     r (1 - s) \left(D - s - \frac1s \right) 
\end{eqnarray}
The second copy of the Maxwell field has negative parity and it can be understood as an antifield for the BV formalism. 

In order to better distinguish between fields and antifields, we first provide a rewriting of \eqref{matE} as follows 
\begin{eqnarray}
    \label{matLA}
    \omega_{\text{I}}^{r=1} = 
    \Big(A(x) + C(x) \beta + \phi^\star(x) \gamma\Big) 
    +\Big(A^\star(x) + \phi(x) \beta + 
    C^\star(x) \gamma\Big) c
\end{eqnarray}
where $A,\phi,C$ are the fields (1-form gauge potential, 0-form auxiliary field, 0-form ghost field) and $A^\star, \phi^\star, C^\star$ are the antifields. In order to relate them, we need to show the BV symplectic form. 

Computing the $\star_{BV}$ Hodge dual, we have 
\begin{eqnarray}
\label{BVB}
\omega^{r=1}_{\text{IV}} = \star_{{BV}} \, \omega^{r=1}_{\text{I}} = \Big(\star A + \star C \partial_\gamma + \star \phi^\star \partial_\beta \Big) \delta(\gamma) \delta(\beta) + 
c  \Big(\star A^\star +  
\star \phi \partial_\gamma + \star C^\star \partial_\beta \Big) \delta(\gamma) \delta(\beta)
\end{eqnarray}
where the derivatives are acting on $\delta(\gamma) \delta(\beta)$ and $\star \Phi$ denotes the space-time Hodge dual of the field $\Phi$, and it is due to $\star_{{BV}}$ acting on $\psi \cong d x$. 

In terms of these ingredients, we can write the BV symplectic structure as follows 
\begin{eqnarray}
\label{BVC}
\Omega_{BV}&=& \langle \omega^{r=1}_{\text{I}}, \omega^{r=1}_{\text{I}} \rangle = \int d^4 xd\gamma d\beta dc d^4\psi \left( \star_{BV} \omega^{r=1}_{\text{I}} 
 \omega^{r=1}_{\text{I}} \right)
\nonumber \\
&=&\int  \Big(\star   A^{\star}  \wedge A + \star C^{\star}\wedge C + \star \phi^{\star}  \wedge  \phi \Big) \, .
\end{eqnarray}

\vspace{0.5cm}

{Let us now go back to our inspection for generic $r$.} The general equations are obtained by acting with $Q$ on a generic representative 
\begin{eqnarray}
\label{cohoB}
Q \omega^r_{\text{I}} &=& \sum_{p,q} c \partial^2 A^{r-p-q}_{q,p}  \gamma^{p} \beta^q 
+ \sum_{p,q} (d^\dagger A^{r-p-q}_{p,q} + 
c d^\dagger \chi^{r-p-q}_{q,p})   \gamma^{q+1} \beta^p \nonumber \\
&+& \sum_{p,q} p\, (d A^{r-p-q}_{q,p} + c d \chi^{r-p-q}_{q,p}) \gamma^q \beta^{p-1} 
+ \sum_{p,q} p\, \chi_{q,p}  \gamma^{q+1}  \beta^{p-1} 
\end{eqnarray}
Being in $\ker Q$ returns two equations 
\begin{eqnarray}
\label{cohoC}
&&\partial^2 A^{r-p-q}_{q,p} + d^\dagger \chi^{r-p-q+1}_{q-1,p} + (p+1) d 
\chi^{r-p-q-1}_{q,p+1}=0\,, ~~~~~\nonumber \\
&& d^\dagger A^{r-p-q}_{q,p} + (p+1) d A_{q+1,p+1}^{r-p-q-2} + (p+1) \chi^{r-p-q-1}_{q,p+1}=0\,. \label{complete-coho}
\end{eqnarray}
Acting with $d$ and with $d^\dagger$ on the second equation we get 
\begin{eqnarray}
\label{cohoD}
 d d^\dagger A^{r-p-q}_{q,p} + (p+1) d \chi^{r-p-q-1}_{q,p+1} =0 \,, \hspace{2cm} d^\dagger d A^{r-p-q}_{q,p} + d^\dagger \chi^{r-p-q+1}_{q-1,p} =0\,. 
\end{eqnarray}
and their sum gives the first eq. (\ref{cohoC}). The solution of these equations can be obtained 
as seen above for the particular case of $r=1$, producing the spectrum provided by the partition function, {namely (two copies of) Maxwell theory for the top $r$-form.}

\vspace{0.5cm}

In order to check the cohomology in the $2$-picture sector, we consider the Hodge dual of \eqref{matB} 
\begin{eqnarray}
    \label{matM}
    \star_{BV} \omega^{r=0}_{\text{I}} = 
    \omega^{r=2}_{\text{IV}} = 
    (A^4_{0,0} + c \chi^4_{0,0}) \delta(\beta) \delta(\gamma) 
\end{eqnarray}
where $A^4_{0,0}, 
\chi^4_{0,0}$ are 4-forms. 
Notice that the Hodge dual is extended also to the spacetime fields, not only to the ghost fields. This concerns also the form number and it is dimensionality dependent. 

Applying the BRST differential, we get the equations 
\begin{eqnarray}
    \label{matN}
c \partial^2 A^4_{0,0} \delta(\beta) \delta(\gamma) + 
d A^4_{0,0} \delta'(\beta) \delta(\gamma) - c 
d \chi^4_{0,0} \delta'(\beta)\delta(\gamma) =0 
\end{eqnarray}
Notice that, since $A^4_{0,0}, 
\chi^4_{0,0}$ are $4$-forms, the last two pieces drop out. The remaining equation 
tells us that $A^4_{0,0}$ 
satisfies the Klein-Gordon equation. As above, by computing the gauge symmetry we see that 
\begin{eqnarray}
\label{mathO}
    \delta \chi^4_{0,0} = 
    \partial^2 \Sigma^4_{0,0}
\end{eqnarray}
which implies that, unless $\chi^4_{0,0}$ is in the kernel of Klein-Gordon operator, it can be removed by that gauge symmetry. Again, there are two degrees of freedom with opposite parity. {As expected, the cohomology in picture $2$ (and dual $r$-charge 2) coincides with what seen for picture $1$, at $r=0$.}

Similarly, starting from $r=1$ $2$-picture case
\begin{eqnarray}
    \label{matP}
    \star_{BV} \omega_{\text{I}}^{r=1} &=& 
    \omega^{r=1}_{\text{IV}} \nonumber \\&=& 
    \Big(A^3_{0,0} + A^{4}_{1,0} \partial_\beta + A^4_{0,1} \partial_\gamma\Big) \delta(\beta) \delta(\gamma) + 
    c \Big(\chi^3_{0,0} + \chi^{4}_{1,0} \partial_\beta + 
    \chi^4_{0,1} \partial_\gamma\Big) \delta(\beta) \delta(\gamma)
 \end{eqnarray}
by the same techniques, it can be shown (not easily) that the $3$-forms 
$A^3_{0,0},\chi^3_{0,0}$, satisfy Maxwell equation. This proves that indeed, there is a duplication of the cohomology both in the superform sector $\omega_{\text{I}}$ and in the integral form sector 
$\omega_{\text{IV}}$. 

Again, for generic $r>1$ charge, we can write the general equations
 \begin{eqnarray}
\label{coho2A}
\omega^r_{\text{IV}} = \sum_{p\geq 0, q\geq0} (A_{p,q} + \chi_{p,q}^* c) \delta^{(p)}(\gamma) \delta^{(q)}(\beta) 
\end{eqnarray}
Acting with BRST charge $Q$, we get 
\begin{eqnarray}
\label{coho2B}
Q \omega^r_{\text{IV}} &=& \sum_{p, q} 
c \partial^2 A_{q,p}\delta^{(p)}(\gamma) \delta^{(q)}(\beta) + 
\sum_{p,q} (-q) (d^\dagger A_{q,p} + 
d^\dagger \chi_{q,p} c) \delta^{(q-1)}(\gamma) \delta^{(p)}(\beta) 
\nonumber \\
&+& \sum_{p,q}  (d A_{q,p} + d \chi_{q,p} c) \delta^{(q)}(\gamma) \delta^{(p+1)}(\beta) +
\sum_{p,q} (-q) \chi_{q,p}  \delta^{(q-1)}(\gamma) \delta^{(p+1)}(\beta)
\end{eqnarray}
and this gives 
\begin{eqnarray}
\label{coho2B}
&&\partial^2 A_{q,p} - (q+1)d^\dagger \chi_{q+1,p} + d \chi_{q,p-1}=0\,, ~~~~~\nonumber \\
&&- (q+1) d^\dagger A_{q+1,p} + d A_{q,p-1} - (q+1) \chi_{q+1,p-1}=0\,. 
\end{eqnarray}
Once again, the antifields may have a $d$-exact and a co-exact part, depending on the R-charge. The R-charge dictates also whether all the closure, co-closure conditions survive or not.

\subsection{Spectral Sequence}

The logic of section \ref{sec:BRST-coh} was to proceed in layers, starting with the cohomology of pure ghost $Q_0$ and adding extra ingredients later to obtain the full  cohomology of $Q$.
In this section we will use the logic of  spectral sequences(see e.g. \cite{Bott:1982xhp}) to deal with these two differential operators.  

What we have observed in the previous section is that the twisted partition function for the ghost system ($Q_0$ cohomology) differs from the computation of the twisted partition function computed in section \ref{section3} by the ghost $c$ enumerating factor. Now we will rectify this mismatch.

We first observe that the partition function for $\mathcal{N}=2$ case is
\begin{eqnarray}
    \label{sepA}
    {\rm T} \mathbb{P}^{N=2}(r,s) = \frac{(1-s) (1 - r)^D}{(1-  r s)\left(1 - \frac{r}{s}\right)}
\end{eqnarray}
and the partition function for the ghost system is 
\begin{eqnarray}
    \label{sepB}
    {\rm T}\mathbb{P}^{\text{ghost}}(r,s) = \frac{(1-s^2) }{(1-  r s)\left(1 - \frac{r}{s}\right)}
\end{eqnarray}
Notice that the ghost factor $(1-s^2)$ is different from the two partition functions. In order to relate them we can multiply and divide $ \mathbb{TP}^{N=2}(r,s)$ by $(1-s^2)$ to get
\begin{eqnarray}
    \label{sepC}
    {\rm T}\mathbb{P}^{N=2}(q ,r,s) = \frac{(1-s) (1-s^2)(1 -  r)^D}{(1-s^2)(1-  r s)\left(1 - \frac{r}{s}\right)} = 
    \left( \frac{(1-s)(1 -  r)^D}{(1-s^2)}  \right) {\rm T}\mathbb{P}^{\text{ghost}}(r,s)
\end{eqnarray}
where we have separated the ghost function from the entire expression. Therefore, the computation of the spectral sequence should give us the factor 
${(1-s)(1 - r)^D}/{(1-s^2)}$. 
The introduction of the $(1-s^2)$ in the numerator 
implies that there should be a new anticommuting field $w$. However, introducing a new field {on its own} would spoil the cohomology computation unless {it is accompanied by} 
a BRST partner $u$ which has the same degree 
but is a commuting field. The full BRST operator is thus modified as follows 
\begin{eqnarray}
    \label{sepD}
    Q \rightarrow Q + w \frac{\partial}{\partial u} = 
    c P^2 + \gamma \bar\psi\cdot P + \psi\cdot P 
    \frac{\partial}{\partial 
\beta} - \gamma \frac{\partial}{\partial \beta}\frac{\partial}{\partial c}  + w \frac{\partial}{\partial u}
\end{eqnarray}
The last piece contains the BRST doubles $w, u$ and 
this ensures that the cohomology is indeed independent from $u,w$. Nonetheless, it is useful because we can construct two independent BRST charges as follows: we can simply redefine the fields $c = c', w = w' + c'$ to get
\begin{eqnarray}
    \label{sepE}
Q = c' P^2 + \gamma \bar\psi\cdot P + \psi\cdot P 
    \frac{\partial}{\partial 
\beta} - \gamma 
\frac{\partial}{\partial \beta} 
\left(\frac{\partial}{\partial c'} - \frac{\partial}{\partial w'}\right)  + (w' + c') \frac{\partial}{\partial u}  
\end{eqnarray}
and finally we can separate into the two pieces
\begin{eqnarray}
    \label{sepF}
    Q_K &=& 
    c' P^2 + \gamma \bar\psi\cdot P + \psi\cdot P 
    \frac{\partial}{\partial 
\beta} - \gamma 
\frac{\partial}{\partial \beta} \frac{\partial}{\partial c'}   + w' \frac{\partial}{\partial u} \,,  
\\
Q_C &=& \gamma 
\frac{\partial}{\partial \beta} \frac{\partial}{\partial w'} + c'  \frac{\partial}{\partial u} \,,
\end{eqnarray}
which satisfy
\begin{eqnarray}
    \label{sepG}
    Q_K^2 =0\,, \hspace{1.5cm}
    \{Q_K, Q_C\} =0\,, \hspace{1.5cm}
    Q^2_C =0\,. 
\end{eqnarray}
Note that the cohomology of $Q_C$ is identical to that computed above for $Q_0$, where the role of $c$ is replaced by $w$. In addition, the second term implies that the cohomology does not depend upon $c'$ and $u$. 
Finally, the spectral sequence implies that 
\begin{eqnarray}
\label{sepI}
 \mathbb{H}_{Q}(V) = \mathbb{H}_{Q_K}( \mathbb{H}_{Q_C}(V)) \,,
\end{eqnarray}
where $V$ is an off-shell Fock space.
This confirms that the computation done in the previous sections can also be done using the methods of spectral sequences.

At last, we should discuss the weight assignment for the new fields $u,w$. It is convenient to assign $s$-degree equal to $+2$ for $w$ and $+2$ for 
$u$. This leads exactly to the partition function $(1-s^2)/(1-s^2) =1$, which means that this is a BRST doublet. According to this grading the two BRST charges 
decompose as follows 
\begin{eqnarray}
    \label{sepL}
    Q_K = Q_{K,1} + Q_{K,0}\,, \hspace{1.5cm} 
    Q_C = Q_{C,-1} + Q_{C,0}\,,
\end{eqnarray}
According to this grading we see that $Q_{C,-1}^2 = Q_{C,0}^2 = \{Q_{C,-1}, Q_{C,0}\} =0$ and 
$Q_{K,1}^2 = Q_{K,0}^2 = \{Q_{K,1}, Q_{K,0}\} =0$
and finally 
$\{Q_{K,1}, Q_{C,-1}\} + \{Q_{K,0}, Q_{C,0}\} =0\,.$



\section{$\mathcal{N}=2$ Action}
\label{sec:action}
The previous sections were concerned with cohomology of world line field theory with the equivalence relations reflecting the linear gauge redundancies of the dynamical system. In the absence of further data there is not a unique non-linear extension (if any) of the linear gauge system. The standard (Noether) procedure to construct a non-linear extension is to find a product $m_2$ on the space of world line fields that commutes with BRST operator $Q$. This $m_2$ then defines the quadratic extension of the linear gauge transformation. Then one proceeds recursively for the higher interaction terms. In this section, we construct an action for $R=1$ subspace of $V$ in $D=4$, which reproduces Yang-Mills theory. It features  essentially three pieces that have to be constructed: the term quadratic 
in the quantum fields, the trilinear terms (which usually follow {\it almost} 
naturally from the quadratic part of the theory) and, eventually, the quadrilinear part. The closure of the gauge algebra implied by the interaction  terms is crucial for gauge symmetry. In the case of Yang--Mills Lie algebra, the action truncates at the fourth order. In the case of higher symmetries, the action does not truncate to 
fourth order, but this fact will not be covered here (see \cite{Jurco:2018sby, Bonezzi:2022bse}). 

\subsection{Quadratic part}

Recall the general state in picture zero and $R=1$ is
\begin{equation}\label{QA}
    \omega^{(0)}  = \left(A(x) + C(x) \beta + \phi^\star(x) \gamma \right)  + \left(A^\star(x) + \phi(x) \beta + 
    C^\star(x) \gamma\right) c
\end{equation}
where we used a simplified notation for the component fields $A(x), C(x), \phi(x)$ and their antifields $A^\star, C^\star, \phi^\star$. 
The ghost assignment can be done consistently with the worldline ghost number by requiring that $\omega^{(0)}$ 
has total (target space plus worldline) ghost number equal to zero. 

This action can written as an integral form built out of the $0$-picture state  $\omega^{(0)}$  and 
 the $2$-picture $\omega^{(2)}$ built by using the Hodge dual operator {$\star_{BV}$}. Notice that since we have established that the free equations are $Q \omega^{(0)} =0 $, it is natural to construct an action, however to avoid the duplication of degrees of freedom, we set 
  $\omega^{(2)} =\star_{BV} \omega^{(0)}$.

  For that we observe that there are two cohomology classes $\mu$ and $\mu'$ (they differ by the statistics and by the ghost number) that can be used to define a  measure for the action by setting $\langle \mu \rangle =1$, where $\langle \cdot \rangle$ denotes the integration as in \eqref{eq:S_M}. Here we use the measure defined by $\mu$ ($\mu'$ will be used in the quartic piece of the action). Notice that inserting the 
  BRST charge commute with $\mu$ and therefore it is compatible. 
Explicitly, we have 
   \begin{align}\label{eq:S_M}
    S[A]=\langle \omega^{(2)} Q\,\omega^{(0)} \rangle 
    = \int d^4 xd\gamma d\beta dc d^4\psi \left( \star_{BV} \omega^{(0)} Q\,\omega^{(0)}\right)\,.
    \end{align}
  Using the identification \eqref{QA} the classical part (found at target space ghost number zero, i.e. $\omega^{(0)}_{\text{ghost}=0} = A(x) + c \beta \phi(x)$) is
\begin{eqnarray}
\label{actA}
S[A] &=& \int \mathrm{d}^4 x \, A^{ \mu} \left( \square A_{\mu} - \partial_\mu \phi \right) + 
\phi \left( \partial^\mu A_\mu - \phi \right) \,  \nonumber \\
&=& \int \mathrm{d}^4 x \, \left(A^\mu \square A_{\mu} 
+ 2  \phi  \partial^\mu A_\mu - ( \phi )^2 \right) \,.
\end{eqnarray}
Eliminating the auxiliary field $ \phi $ we retrieve the gauge invariant Maxwell action, 
$S[A] = -\frac{1}{2} \int \mathrm{d}^4 x F^{\mu\nu} F_{\mu\nu}$. 
The space-time gauge transformations resulting from  $\delta_\Lambda\omega^{(0)}_{\text{ghost}=0} =Q\Lambda$ are given by 
\begin{eqnarray}
\label{actB}
\delta A = d \lambda\,, ~~~~~~
\delta  \phi  = \square \lambda \, .
\end{eqnarray}
Completing the computation of all terms, including the antifields in 
\eqref{eq:S_M} we get the full BV action.  $C(x)$ is a scalar field of ghost degree $+1$, while $A^\ast$ and $\phi^{\ast}$ have ghost number $-1$, and $C^\ast$ has ghost number $-2$ and it represents the 
antifield to a target space ghost $C$. 
The antifield contribution of the action \eqref{eq:S_M} reads
\begin{equation}
    S_{\text{af}} = 2 \int \mathrm{d}^4 x \, 
    \left( A^{*}_\mu \partial^\mu C(x) - A^{*} \square C(x) \right)\,, 
\end{equation}
as expected for a quadratic theory with an auxiliary field. Notice that $C^\ast$ does not appear at quadratic order. 

Given the BV action, we can easily obtain the full BRST-BV symmetry. We denote by ${\mathfrak s}_0$ the BRST-BV symmetry restricted to the quadratic level for the action (at linear level for the transformation rules) we can get the transformation rules by denoting by $\omega(\Phi)$ the expression \eqref{QA} in terms of the target space fields $\Phi = C, A, \phi, C^*, A^*, \phi^*$ and 
setting 
\begin{eqnarray}
    \label{poffaA}
    \omega({\mathfrak s}_0 \Phi) = Q  \omega(\Phi)\,,
\end{eqnarray}
where $Q$ acts on the worldline fields which $s_0$ acts on the target space fields. From this equation, we get the BRST-BV transformation rules
\begin{eqnarray}
\label{BVE}
 &&{\mathfrak s}_0\, A = d C\,, ~~~~~~~
 {\mathfrak s}_0 \,C = 0\,,  ~~~~~~
 {\mathfrak s}_0\, \phi = \star d \star d  C = d^\dagger d C \,, 
 \nonumber \\
 &&{\mathfrak s}_0\, A^{\star} = (d d^\dagger + d^\dagger d) A  + d \phi\,, ~~~~~~~
 {\mathfrak s}_0\, C^{\star} = d^\dagger d \phi^{\star} + d^\dagger A^{\star}\,, ~~~~~~~
 {\mathfrak s}_0\, \phi^{\star} = d^\dagger  A + \phi\,.
 \end{eqnarray}
Note that the first equation in the second line can be rewritten as 
${\mathfrak s}_0\,(A^{\star} + d \phi^{\star} ) = d^\dagger F$ where 
$ F = dA$ is the usual field strength. It is easy to check the nilpotency of ${\mathfrak s}_0$. Introducing the new antifield $\widetilde A^{\star}  = (A^{\star} + d \phi^{\star} )$, we can rewrite the BRST-BV transformations 
as follows
\begin{eqnarray}
\label{BVF}
 &&{\mathfrak s}_0\, A = d C\,, ~~~~~~~
 {\mathfrak s}_0 \,C = 0\,,  ~~~~~~
 {\mathfrak s}_0\, \phi  = d^\dagger d C \,, 
 \nonumber \\
 &&{\mathfrak s}_0\,  \widetilde A^{\star} = d^\dagger F \,, ~~~~~~~
 {\mathfrak s}_0\, C^{\star} = d^\dagger \widetilde A^{\star}\,, ~~~~~~~
 {\mathfrak s}_0\, \phi^{\star} = d^\dagger  A + \phi\,.
 \end{eqnarray}

In the same way, we can start from another picture to get the same result. Since the PCO operators commute 
with the BRST charge, we can use them to change the picture of the vertex operator. { We have noticed  in sec. \ref{sec:BRST-coh} that 
the complete map between zero picture and one picture can be achieved by combining the picture changing operator $Y(\gamma)$ and $Y(\beta)$ (defined in  
\eqref{PCO:increase-pic}). This leads to 
$\omega^{(0,1)} = (Y(\beta) + Y(\gamma))\omega^{(0)}$, and therefore we can construct the action in the other picture as 
  \begin{eqnarray}
\label{BVH}
  S[A]= & \int d^4 xd\gamma d\beta dc d^4\psi \left( \star \left( (Y(\beta) + Y(\gamma)) \omega^{(0)}\right)  Q\,(Y(\beta) + Y(\gamma)) \omega^{(0)}\ \right) 
\end{eqnarray}
and using the Hodge dual operation, the 
vertex $\omega^{(0,1)}$ is mapped into a complementary pictured space and we can reproduce the original action. }
  
%

\subsection{Trilinear terms}
We now seek a 2-product of world line fields that reproduces the quadratic gauge transformation of Yang-Mills theory\footnote{Here we do not explore the possibility of other extensions. } by promoting ordinary space-time  derivatives to covariant derivatives. For that purpose, we first 
introduce new differentials $D$ and $D^\dagger$ forming  the algebra
\begin{eqnarray}
    \label{PC}
    &&D = \Big(d + \gamma \frac{\partial}{\partial c}\Big)\,, \hspace{1.5cm}
    D^\dagger = \left( d^\dagger + \frac{\partial}{\partial c} \frac{\partial}{\partial \beta}\right) \nonumber \\
    && D^2 = (D^\dagger)^2 = 0 \,, \hspace{1.5cm} 
    D \, D^\dagger + D^\dagger \, D = - \partial^2\,.
\end{eqnarray}
Those generalize the differential operator $d, d^\dagger$ in the presence  ghost fields $\beta$ and $\gamma$. The two differential 
operators commute with the BRST operator. To prove that,  we observe that 
\begin{eqnarray}
    \label{commuA}
    D = [Q, \beta]\,, ~~~~~~
    D^\dagger = \left[ \frac{\partial}{\partial \gamma},Q \right]
\end{eqnarray}
from which it follows that $[Q,D]= [Q, D^\dagger]=0$, using the Jacobi identity.  Another crucial relation is 
\begin{eqnarray}
    \label{commuB}
    Q = D c D^\dagger + D^\dagger c D -  Q_0\,.
\end{eqnarray}
Using the BV Hodge dual $\star_{BV}$, we get 
\begin{eqnarray}
    \label{commuC}
    \star_{BV} D \star_{BV} = D^\dagger\,, ~~~~~
    \star_{BV} D^\dagger \star_{BV} = D\,
\end{eqnarray}
that can be checked easily. 
Let us furthermore introduce two new differential operators 
\begin{eqnarray}
    \label{addA}
    \Sigma &=& [Q, c \beta ] = 
    - c D + \gamma \left(1 + \beta
    \frac{\partial}{\partial \beta} \right) \nonumber \\
    \Sigma^\dagger &=& \left[Q, c\frac{\partial}{\partial \gamma}\right] = 
    c D^\dagger + \gamma 
    \frac{\partial}{\partial \gamma} 
    \frac{\partial}{\partial \beta} \,,
\end{eqnarray}
with
\begin{eqnarray}
    \label{addB}
    \star_{BV} \Sigma \star_{BV} = \Sigma^\dagger\,.
\end{eqnarray}
Notice the need of the transposition. The commutation relation between $\Sigma$'s and $D$'s are 
\begin{eqnarray}
    \label{addC}
    [D, \Sigma^\dagger] &=& c \partial^2 
    + \gamma d^\dagger\,, \hspace{1cm} 
    [\Sigma,D^\dagger ] = c \partial^2 + d \frac{\partial}{\partial \beta} \,.
\end{eqnarray}
Note that the differential operators $c \partial^2 
    + \gamma d^\dagger$ and $c \partial^2 + d \frac{\partial}{\partial \beta} $ 
    are BRST invariant due to the Jacobi identites. They are also BRST exact. In addition, the relations \eqref{addC} are compatible with $\star_{BV}$. 

Let us now compute the Hodge dual of $c D$ and of $c D^\dagger$. 
\begin{eqnarray}
    \label{addD}
    \star_{BV} (c D) \star_{BV} = D^\dagger c = 
    \left(d^\dagger + \frac{\partial}{\partial \beta} \frac{\partial}{\partial c}\right) c = 
    - c d^\dagger + \frac{\partial}{\partial \beta} \left(1 - c  \frac{\partial}{\partial c}\right) = - c D^\dagger + \frac{\partial}{\partial \beta}\,.
\end{eqnarray}
Therefore it would be preferable to introduce new operators which are BRST invariant as follows  
\begin{eqnarray}
     \label{addH}
      \Gamma^\dagger &\equiv& \Sigma^\dagger - \frac{\partial}{\partial \beta} 
      = 
       c d^\dagger  + \left(c  \frac{\partial}{\partial c} + \gamma  \frac{\partial}{\partial \gamma}-1\right) \frac{\partial}{\partial \beta}
      \,, \nonumber \\
      \Gamma &\equiv& \Sigma - \gamma
      = - c d + \gamma \left( \beta\frac{\partial}{\partial \beta} - c 
       \frac{\partial}{\partial c}\right)
      \,, 
 \end{eqnarray}
from which it  immediately follows that $(\Gamma)^\dagger = \star_{BV} \Gamma 
\star_{BV}= \Gamma^\dagger$. We would like to point out that $ \Gamma^\dagger $ is closed since $\partial_\beta$ does commute with $Q$, but it is not exact unless we {\it escape} in the Large Hilbert Space
\begin{eqnarray}
    \label{addHA}
     \Gamma^\dagger = \left\{Q, c\frac{\partial}{\partial \gamma} - \frac{c}{\gamma}\right\}\,,
\end{eqnarray}
where $c/\gamma$ lives in the LHS. Note that the brackets in $\Gamma^\dagger$ and $\Gamma$ acting on polynomials of fields have the following kernels 
\begin{eqnarray}
    \label{addHB}
    {\rm ker} \left(c  \frac{\partial}{\partial c} + \gamma  \frac{\partial}{\partial \gamma}-1\right) = \{c, \gamma\} \nonumber \\
    {\rm ker} \left( \beta\frac{\partial}{\partial \beta} - c 
       \frac{\partial}{\partial c}\right) = 
       \{{\rm constants}\} \,.
\end{eqnarray}
The expression $\Gamma^\dagger$ is crucial to get the correct gauge transformations. First, we note that 
\begin{eqnarray}
    \label{addHC}
    \Gamma^\dagger \omega^{R=1} = \left(
       c d^\dagger  + \left(c  \frac{\partial}{\partial c} + \gamma  \frac{\partial}{\partial \gamma}-1\right) \frac{\partial}{\partial \beta} 
       \right) 
       ( A + c \beta \phi)  = c d^\dagger A
\end{eqnarray}
while the rest drops out. 

The introduction of these new differential operators is important to built the non-linear gauge transformations and to reproduce the usual Yang-Mills gauge transformations to second order. For instance, restricting to the physical field expression 
$\omega^{R=1} = A + c\beta \phi$ (setting to 
zero the antifields) and for a gauge parameter is $\Lambda^{R=1} = \beta \lambda$, 
we set 
\begin{eqnarray}
    \label{addI}
    \delta \omega^{R=1} &=&Q( \Lambda^{R=1}) + \Gamma^\dagger [\omega^{R=1}, \Lambda^{R=1}] \quad\implies\nonumber \\
   \delta A + c \beta \delta \phi &=& ( c \beta \partial^2 \lambda + d \lambda) + 
    \left(c d^\dagger  + (c  \frac{\partial}{\partial c} + \gamma  \frac{\partial}{\partial \gamma} -1) \frac{\partial}{\partial \beta} \right) (\beta [A, \lambda] + \beta^2 c [\phi, \lambda]) \nonumber \\
    &=& ( c \beta \partial^2 \lambda + d \lambda) + c \beta d^\dagger [A,\lambda] {-} [A,\lambda]\,,
\end{eqnarray}
where the second equation is obtained by expanding all terms defined in the first line. Thus,
\begin{eqnarray}
    \label{addL}
    \delta A &=& d \lambda {-} [A, \lambda]\nonumber \\
    \delta \phi &=& {-} d^\dagger \left( d \lambda + [A, \lambda] \right)\,.
\end{eqnarray}
Once that the non-linear terms are constructed, we can try to recast them into a cyclic product as in  \cite{Jurco:2018sby} by setting  
\begin{eqnarray}
    \label{addM}
m_2({\mathcal A}, {\mathcal B} ) = \Gamma^\dagger 
\left[{\mathcal A}, {\mathcal B} \right] + \star_{BV}\left[{\mathcal A}, \Gamma^\dagger \star_{BV}{\mathcal B} \right] + \star_{BV}\left[\Gamma^\dagger \star_{BV} {\mathcal A}, {\mathcal B} \right] \,,
\end{eqnarray}
where $\left[{\mathcal A}, {\mathcal B} \right]$ is the Lie algebra bracket. First of all, we notice that the first terms is what was constructed in \eqref{addI}. The other terms can be inferred from cyclicity as follows. 
Let us consider the action
\begin{eqnarray}
    \label{addMAA}
   S_3  =  \langle \star_{BV} \omega \Gamma^+ [\omega, \omega] \rangle\,,
\end{eqnarray}
where we have inserted the non-linear terms of \eqref{addI} and we performed a variation with respect to $\omega$, giving 
\begin{eqnarray}
    \label{addMMAB}
    \delta S_3 &=& \langle \star_{BV} \delta \omega \Gamma^\dagger[\omega, \omega] + 
    \star_{BV} \omega \Gamma^\dagger[\delta \omega, \omega] + 
    \star_{BV} \omega \Gamma^\dagger[\omega, \delta\omega]\rangle \nonumber \\
    &=&\langle \star_{BV} \delta\omega \left( 
    \Gamma^\dagger [\omega, \omega] + \star_{BV} [\Gamma^\dagger \star_{BV} \omega, \omega] 
    + \star_{BV} [\omega, \Gamma^\dagger \star_{BV} \omega] \right) 
\rangle \,.
\end{eqnarray}
Then bringing $\delta \omega$ in evidence in each single piece 
we get \eqref{addM}. The same strategy has been pursued in  \cite{Jurco:2018sby}. Note that this 2-bracket is compatible with the 
BRST charge because $Q$ commutes with $\Sigma^\dagger$ and with the Hodge dual operator $\star_{BV}$, i.e. 
\begin{eqnarray}
    \label{addMA}
    Q m_2({\mathcal A}, {\mathcal B} )    &=& m_2( Q{\mathcal A}, {\mathcal B} ) 
    \pm  m_2({\mathcal A}, Q{\mathcal B} )\,.
\end{eqnarray}
Let us then display the full result for the cubic terms
 \begin{eqnarray}
\label{NONA}
S_3  
&=& \int d^4x {\rm Tr}\left(- d A\wedge \star [A,A] - \phi^\star [d \star A, C]\right) + \nonumber \\
&+& \int d^4x {\rm Tr}\left(\star A [ C, A^\star] + \star A [ A, \phi] + \star C [C, C^\star] + \star C [\phi^\star, \phi] + 
2 \star \phi^\star [C, \phi] 
\right) \,,
\end{eqnarray}
which is just what is required for invariance under all non-abelian BRST transformations, up to second order, of the fields in the spectrum. 
For example, by taking the derivatives with respect to $A^\star$ or $\phi^\star$ or $C^\star$, 
one obtains the variation of $A, \phi$ and $C$, while taking the variation with respect to $A, C, \phi$, 
one gets the BRST variation of the antifields and therefore also the equations of motion. They read 
at this level as 
\begin{align}
\label{BVG}
 {\mathfrak s}_1\, A &= d c + [A,c]\,,  \nonumber \\
 {\mathfrak s}_1\, \hat \phi &= d \star d c +d \star [A, c] + [(A), \star d c] +[(A), \star [A, c]] + [ d\star A, c] +[\hat\phi, c]\,,  \nonumber \\
 {\mathfrak s}_1\, c &= \frac12 [c,c]\,,   \nonumber \\
 {\mathfrak s}_1\, \hat A^{\star} &= d \star d A + d\star \frac12 [A,A] + [A,  \star d A]   - [\hat A^{\star},  c] \,, 
 \nonumber \\
 {\mathfrak s}_1\, \hat \phi^{\star} &= d \star A + \hat\phi - [\hat \phi^{\star}, c]\,, \nonumber \\
 {\mathfrak s}_1\, \hat c^{\star} &= d \hat A^{\star} + [A, \hat A^{\star}] + [\hat c^\star, c] + [\hat\phi^{\star},\phi]\,. ~~~~~~~
  \end{align}
Notice that there is a missing term $ \frac12 [A, \star[A,A]]$ in the BRST transformation of the 
antifield $A^\star$ since it appears at higher order. We know that such a term emerges from the consistency of the Yang-Mills algebra and from the non-associativity of $m_2$. In general, we should test the non-associativity of the $m_2$ product to 
produce $m_3$ product as 
\begin{eqnarray}
    \label{m3A}
    m_2(A, m_2(B,C)) &+& m_2(m_2(A,B),C) = Q m_3(A,B,C) \nonumber \\
    &+& m_3(QA, B, C) + m_3(A, QB, C) + m_3(A, B, QC)\,.
\end{eqnarray}
We tested on all {\it target space fields} reproducing the results of  \cite{Jurco:2018sby}, nonetheless we are unable to construct the $m_3$ product for the vector space $V$ of world-line fields as we did for $m_2$ above. As mentioned above, it is not guaranteed that such triple product on $V$ exists. Notice that, since we wish to reproduce the Yang-Mills algebra, we do not expect higher products $m_n$. The $A_\infty$ algebra, if it exist, will be truncated at the third level. 

\subsection{Quartic Term}\label{sec:S4}
Since we cannot complete the quartic term in the action by cancelling the non-associativity of $m_2$, we will complete the construction of the action by requiring invariance under the space-time Yang-Mills-BV transformations. Recall that so far we have used the measure induced by the cohomology class $\mu$ related to the $\star_{BV}$. 
As been discussed above that measure induces the BV symplectic structure, but there is a companion class $\mu'$ which provide the metric structure and in terms of it we can construct the quartic terms. Notice that, it is really the presence of the two cohomology classes $\mu$ and $\mu'$ which allows us to truncate the action to the quartic term.

Let us study the last term, the quadratic one
\begin{eqnarray}
\label{NONf}
S_4 = \Big\langle \star'[\omega, \omega] [\omega, \omega] \Big\rangle
\end{eqnarray}
where we compute the product $[\omega, \omega]$ and then we act with the 
$\star'$ operation on it, and we take the wedge product with $[\omega, \omega]$.
{ The $\star'$  is defined as follows 
\begin{eqnarray}
    \label{starprime}
    \star' = \star_{BV} \star_c\,, ~~~~~
    \star_c F(c) = \int d\eta e^{\eta c} F(\eta)
\end{eqnarray}
where $\eta$ is an auxiliary anticommuting variable and 
the integral is a Berezin integral. The $\star'$ correspond to the cohomology class $\mu'$ as pointed out in sec. 5.1. }

In the end, we also take the trace over Lie algebra indices. Notice that, since the $\star'$ has already the 
$c$ ghost in the formula, we are left with only the contributions from $ \omega$ 
independent of $c$, therefore we use the formula above \eqref{NONf} to select the contribution 
\begin{eqnarray}
\label{NONg}
S_4 &=& \int d^4x {\rm Tr}\Big( (\star [A,A] +\star [A,C] \partial_\gamma + \star [A, \phi^\star]\partial_\beta) \delta(\gamma)\delta(\beta) \wedge ([A,A] + \beta [A,C] +  \gamma [A, \phi^\star]) \Big) \nonumber \\
&=&  \int d^4x {\rm Tr}\Big( (\star [A,A]) [A,A] + \star[A,C] [A, \phi^\star] + \star [A,\phi^\star][A, C]\Big)  
\end{eqnarray}
using the fact that $\star[A,C] = [\star A, C]$ since $C$ is a $0$-form on the worldvolume, using the 
Jacobi identities for the Lie algebra brackets, and $[\star A, A]= [A_\mu, A^\mu] {\rm Vol}_4 =0$ 
where ${\rm Vol}_4$ is the volume of the worldvolume. Therefore the last two terms drop out.         
Finally, we can provide the full Yang-Mills-BV transformations  
as 
\begin{align}
\label{BVF}
 {\mathfrak s}\, A &= d_A c \,,  \nonumber \\
 {\mathfrak s}\, \hat \phi &= d_{(A)} \star d_A  c + [ d\star A, c] +[\hat\phi, c]\,,  \nonumber \\
 {\mathfrak s}\, c &= \frac12 [c,c]\,,   \nonumber \\
 {\mathfrak s}\, \hat A^{\star} &= d_A \star F  - [\hat A^{\star},  c] \,, 
 \nonumber \\
 {\mathfrak s}\, \hat \phi^{\star} &= d \star A + \hat\phi - [\hat \phi^{\star}, c]\,, \nonumber \\
 {\mathfrak s}\, \hat c^{\star} &= d_A \hat A^{\star} + [\hat c^\star, c] + [\hat\phi^{\star},\phi] \,, ~~~~~~~
  \end{align}
where $d_A \star A = d\star A$. 
As explained above, in the case of Yang-Mills, we are unable to provide the full algebra on the vector space of world-line fields but we can reproduce the Yang-Mills algebra at the level of the target space fields. The construction of the former for Yang-Mills theory, or higher forms, is an interesting open problem.  

\subsection{Presymplectic Approach}
\label{ivoac}
In this section we formulate a non-linear generalization of the space-time action \eqref{eq:S_M} that is based on the operator algebra rather that world-line fields (or wave functions). In this approach one considers a suitable space of ghost degree one operators $Q(\Phi)$ on $V$, where $\Phi\in \mathcal{F}$, where $\mathcal{F}$ is the linear space-time fields parametrizing $Q$ (see \cite{Grigoriev:2021bes} for more detailed discussion).  The map 
\begin{align}
 Q(\cdot)  &:\mathcal{F\to} V   \end{align}
is surjective so that there is an element in $\mathcal{F}$ of every element in $V$. However, it fails to be injective which means there are {\it background fields} that are not parametrized by elements in $V$.  Concretely, we parametrize deformations of $Q$ by
\begin{align}\label{eq:Qfi}
   Q(\Phi)=&-c\left(p^2+ p\cdot B+B\cdot p
   -G_{\mu\nu}\psi^\mu\bar\psi^\nu
   -\tilde\phi\right)+\gamma \Pi\cdot\bar\psi+\bar\gamma \Pi\cdot\psi+C\\&
    -c\bar\gamma\psi\cdot A^{\star}
   +c\gamma \bar\psi\cdot A^{\star}+\gamma\bar\gamma\phi^{\star}+c\gamma\bar\gamma C^{\star} + \gamma\bar\gamma b\,,
\end{align}
where $\Pi_\mu=p_\mu+A_\mu$ and $\tilde\phi=\phi+[p,B]$. 
Then
\begin{align}\label{eq:sur_1}
 Q(\cdot) &:\mathcal{F\to} \mathcal{H}\nonumber\\
   \Phi&\mapsto Q(\Phi)\beta= \Big(A_{\mu}\psi^\mu + \beta C + \gamma \phi^\star\Big) + 
c  \Big(A^{\star}_{\mu}\psi^\mu + \beta \phi + \gamma C^{\star}\Big)=\omega^{(0)}\,,
   \end{align}
defines a surjective linear 
map form the space-time fields to the space of states of the world line at R-charge 1.\footnote{Note that the position of the $c$-ghost is immaterial since the fields in the bracket have even Grassmann parity.}  Note that there is an ambiguity in the choice of $Q(\Phi)$ in the terms that have vanishing action on $\beta$. Concretely, the fields $B$ and $G_{\mu\nu}$ are not fixed by the surjective map \eqref{eq:sur_1}. One (non-canonical) way to fix this ambiguity is to require a given non-linear BV-transformation of the space-time fields. For this we  recall that \cite{Grigoriev:2021bes}
\begin{align}\label{eq:bvs}
\Omega: Q(\cdot)\mapsto \frac{1}{2}[Q(\cdot),Q(\cdot)]\,,
    \end{align}
 induces a BV-vector field $\mathfrak{s}$ on $\mathcal{F}$ whose components are given by the components of $[Q(\Phi),Q(\Phi)]$. Concretely, 
\begin{align}\label{eq:vfq2}
\frac{1}{2}[Q(\omega),Q(\omega)]=&[Q(\omega),C]+\frac{1}{2}C^2\\&-c\bar\gamma\psi^\mu\left([p^2+p\cdot B+B\cdot p,\Pi_\mu]-G_{\mu\alpha}\Pi^\alpha 
-[\tilde\phi,\Pi_\mu]\right)\nonumber\\&-c\gamma\bar\psi^\mu\left([p^2+B\cdot p+p\cdot B,\Pi_\mu]+\Pi^\alpha G_{\alpha\mu}
-[\tilde\phi,\Pi_\mu]\right) \nonumber \\
\, & -c\gamma\bar\gamma [\Pi^\mu,A^\star_\mu]-c\gamma\bar\gamma[p^2+B\cdot p+p\cdot B-\tilde\phi,\phi^{\star}]+\bar\gamma\gamma (A^2+\tilde\phi)\,,
\end{align}
where $\omega:=\omega^{(0)}$ and we suppressed terms that vanish trivially on the R-charge 1 subspace of $V$. The first line amounts to the adjoint action of $C$ on $\{\Pi, G, C,\cdots\}$. For $\tilde\phi=-A^2$ this reproduces the expected Yang-Mills BRST transformation of the gauge boson $A$. 
If we furthermore set $B=A$ and $G_{\mu\nu}=-2[\Pi_\mu,\Pi_\nu]$ then the second on third line reproduce the expected tranformation of the antifield. 
The choice we made for  $B$ and $G_{\mu\nu}$ was motivated by reproducing the standard BV-transformations for Yang-Mills. To continue we will, however, consider $B$, $\phi$ and $G_{\mu\nu}$ as independent fields. Moreover, $B$ and $G_{\mu\nu}$ should be considered as background fields which don't vary with $\omega$ since they are not determined by the surjection \eqref{eq:sur_1}. We then consider the cubic action
\begin{align}
    S[\omega]=\frac{1}{2}\langle\star_{BV}\left(Q_1(\omega)\beta\right)Q_0Q_1(\omega)\beta\rangle+\frac{1}{3} \langle\star_{BV}\left(Q_1(\omega)\beta\right)Q_1(\omega)Q_1(\omega)\beta\rangle\,.
    \label{qub-BV_action}
\end{align}
The quadratic term  functional $S[\omega]$ is  invariant under the linear BRST transformation transformation 
\begin{align}\label{eq:m1}
    \omega\mapsto m_1(\omega)=[Q_0,Q_1(\omega)]{\beta}\,.
\end{align}
We note, in passing, that for $B_\mu=G_{\mu\nu}=0$, we can furthermore define a bilinear map 
\begin{align}  (\omega_1,\omega_2)\mapsto m_2(\omega_1,\omega_2)&=Q_1(\omega_1)Q_1(\omega_2){\beta}\nonumber\\
&=
\left([\bar\gamma,\omega_1]\omega_2 + \omega_1 [\bar\gamma,\omega_2] + \beta [\bar\gamma,\omega_1][\bar\gamma,\omega_2]\right) 
\end{align}
that is $m_1$-invariant, $m_1\circ m_2 =0$ by \eqref{eq:m1} and associative by the associativity of the algebra of world-line fields. The symplectic form  defined through 
\begin{align}
    <\omega_1, \omega_2>= \langle\star_{BV}\left(Q_1(\omega_1)\beta\right)Q_1(\omega_2){\beta}\rangle
\end{align}
is graded symmetric and $m_2$ is cyclic with respect to it,
\begin{align}
<\omega_1,m_2(\omega_2,\omega_3)>&=\langle\star_{BV}\left(Q_1(\omega_1)\beta\right)(Q_1(\omega_2)Q_1(\omega_3)){\beta}\rangle\nonumber\\
&=\langle\star_{BV}\left(Q_1(\omega_1)\beta\right)Q_1(\omega_2))Q_1(\omega_3){\beta}\rangle\nonumber\\
&=\langle\star_{BV}\left(Q_1(\omega_3)\beta\right)(Q_1(\omega_1)Q_1(\omega_2)){\beta}\rangle\nonumber\\
&=<\omega_3,m_2(\omega_1,\omega_2)>\,.
\end{align}
For $B$ or $G$ different from zero $m_2$ is not well defined. However, we will not need it.  

The quadratic and cubic terms in \eqref{qub-BV_action} are cyclic, provided $B$ and $G_{\mu\nu}$ are considered to be background fields independent of $\omega$. Furthermore \eqref{qub-BV_action} is invariant on the BV-transformation 
\begin{align}
\delta_{BV}Q_1=[Q_0,Q_1]+\frac{1}{2}[Q_1,Q_1]\,,
\end{align}
which is tantamount to the invariance of $S$ under $\Phi\to \mathfrak{s}(\Phi)$ with $ \mathfrak{s}$ defined in \eqref{eq:bvs}. 

Concerning the equations of motions implied by \eqref{qub-BV_action} we apply the variational principle 
\begin{align}\label{qub-BV_action+var}
    \delta S[\omega]&=\langle\star_{BV}\left(Q_1(\delta\omega)\beta\right)[Q_0,Q_1(\omega)]\beta\rangle+ \;\frac{1}{2}\langle\star_{BV}\left(Q_1(\delta\omega)\beta\right)[Q_1(\omega),Q_1(\omega)]{\beta}\rangle\nonumber\\
    &=\frac{1}{2} \langle\star_{BV}\left(Q_1(\delta\omega)\beta\right)[Q_0+Q_1(\omega), Q_0+Q_1(\omega)]{\beta}\rangle\nonumber\\
    &=\frac{1}{2} \langle\star_{BV}\left(Q_1(\delta\omega)\beta\right)[Q(\omega), Q(\omega)]{\beta}\rangle\,
\end{align}
and thus critical points are subject to the equation of motion
\begin{align}\label{eq:pse}
    [Q(\omega), Q(\omega)]{\beta}=0\,.
\end{align}
In \cite{Grigoriev:2021bes} it was shown that with a suitable reality condition the condition $[Q(\omega), Q(\omega)]V_{R=1}=0$ implies the Yang-Mills equations of motion. Eqn. \eqref{eq:pse} is weaker since vanishing is required only when acting on the  constant Yang-Mills ghost state $\beta$  rather than all of $V_1$. 
In component fields, this leads to the equations\footnote{Note that we changed the normalization $B\to -\frac{1}{2}B$ compared to \cite{Grigoriev:2021bes}.}
\begin{align}\label{eq:bfe}
    &\phi+[p,A]+A^2=0\\
    [p^2,A_\mu]+2B^\nu [p_\nu,A_\mu]-&[\phi,\Pi_\mu]-G_{\mu\nu}A^\nu=0\,,\nonumber
\end{align}
where we set the field of odd degree to zero since they cannot be classical. For $B_\mu=A_\mu$ and $G_{\mu\nu}=-2[\Pi_\mu,\Pi_\nu]$ this system reduces to $[\Pi^\mu,[\Pi_\mu,\Pi_\nu]]=0$. Thus the e.o.m.~from \eqref{qub-BV_action+var} are compatible with the Yang-Mills equations. And then the BV-transformations generated by $ \mathfrak{s}$ as defined in \eqref{eq:bvs} are indeed those of Yang-Mills theory. If there are other solutions, they will correspond to alternative BV-extensions of the quadratic Yang-Mills. We note in closing that, for $c$-number valued fields (abelian fields) we can write $B=A+b$ and $G_{\mu\nu}=-2[\Pi_\mu,\Pi_\nu]+g_{\mu\nu}$. Assuming $[\Pi^\mu,[\Pi_\mu,\Pi_\nu]]=0$ we find, upon multiplication with $A^\mu$, that  $b^\nu[p_\nu,A^2]=0$. So, either $A=0$ or $b=0$. In the latter case it follows that $g_{\mu\nu}A^\nu=0$. This suggests that \eqref{eq:bfe} admits no continuously connected deformations of the Maxwell solution.

 \section{Conclusions and outlook}
The purpose of this paper was twofold. Firstly we studied the spectrum of the spinning world line using methods of twisted partition function, and secondly we explored the construction of a string field theory inspired space-time action for world line fields.

The methods of partition functions prove to be particularly effective for extracting coarse information about the spectrum as seen in the examples of this paper. There are multiple interesting examples where the developed techniques can be applied. The easiest application would be to models with extended RNS supersymmetry as partially done in section \ref{ext} and studied before in \cite{Bonezzi:2018box}. These models are natural candidates for first quantized models of higher spin fields. 
A more physically interesting question, albeit difficult, would be to apply the techniques in a world line theory with target space supersymmetry (GS like model or pure spinor string) and compare with results of \cite{Berkovits2018}.

Concerning the target space action side of the story there are several loose ends to be fixed. Although giving rise to a proper target space action, the action presented in \ref{sec:action} has somewhat an ad hoc origin so a more natural construction from the point of view of the world line would be appealing. Furthermore, we believe that ideas put forth in \ref{ivoac} might shed some light on the construction of background independent actions in SFT which we would like to investigate.  Similarly to the previous paragraph any application to SUSY theories of RNS or GS type or chiral models like ambitwistor string to further confirm the validity of these constructions would be welcome.

\section*{Acknowledgement}
 O.H. would like to thank Andres Collinucci for many discussions on the aspects of Hilbert Series. The work of O.\,H.\ was supported by the FWO-Vlaanderen through the project G006119N and by the Vrije Universiteit Brussel through the Strategic Research Program ``High-Energy Physics''.  I.S. is supported by the Excellence Cluster Origins of the DFG under Germany’s Excellence Strategy EXC-2094 390783311 and by grant NSF PHY-2309135 to the Kavli Institute for Theoretical Physics (KITP) I.S. would like to thank the CERN Theory group for hospitality. P. A. G. is partially supported by research funds FAR2019 of  Università del Piemonte Orientale. We would like to thank the Mainz Institute of Theoretical Physics for the workshop "Higher Structures, Gravity and Fields" and the Galileo Galilei Institute in Florence for the workshop "Emergent Geometries from Strings and Quantum Fields" for a great environment which we appreciated during the development of this work.  O.H. would like to thank Universit\`a del Piemonte Orientale for hospitality and financial support during his visit.
 
\bibliographystyle{unsrt}
\bibliography{finaldraft.bib}
\end{document}